\documentclass[journal,10pt,twoside,twocolumn]{IEEEtran}

\usepackage{cite}
\usepackage[T1]{fontenc}
\usepackage{graphicx}
\usepackage{amssymb}
\usepackage{amsmath}
\usepackage{amsfonts}
\usepackage{amsthm}
\usepackage{mathrsfs}
\usepackage{xspace}

\usepackage{paralist}
\usepackage{setspace}
\usepackage{microtype}
\usepackage{hyperref}
\usepackage{url}
\usepackage{balance}
\usepackage{subfigure}
\usepackage{xcolor}
\usepackage{fancyref}
\usepackage{framed}
\usepackage{booktabs}
\usepackage{nicefrac}
\usepackage{textcomp}
\usepackage{multirow}
\usepackage{bm}
\usepackage{upgreek}
\usepackage{stmaryrd}

\usepackage{algorithm}
\usepackage{algpseudocode}

\newcommand{\safemath}[2]{\newcommand{#1}{\ensuremath{#2}\xspace}}

\safemath{\bma}{\mathbf{a}}
\safemath{\bmb}{\mathbf{b}}
\safemath{\bmc}{\mathbf{c}}
\safemath{\bmd}{\mathbf{d}}
\safemath{\bme}{\mathbf{e}}
\safemath{\bmf}{\mathbf{f}}
\safemath{\bmg}{\mathbf{g}}
\safemath{\bmh}{\mathbf{h}}
\safemath{\bmi}{\mathbf{i}}
\safemath{\bmj}{\mathbf{j}}
\safemath{\bmk}{\mathbf{k}}
\safemath{\bml}{\mathbf{l}}
\safemath{\bmm}{\mathbf{m}}
\safemath{\bmn}{\mathbf{n}}
\safemath{\bmo}{\mathbf{o}}
\safemath{\bmp}{\mathbf{p}}
\safemath{\bmq}{\mathbf{q}}
\safemath{\bmr}{\mathbf{r}}
\safemath{\bms}{\mathbf{s}}
\safemath{\bmt}{\mathbf{t}}
\safemath{\bmu}{\mathbf{u}}
\safemath{\bmv}{\mathbf{v}}
\safemath{\bmw}{\mathbf{w}}
\safemath{\bmx}{\mathbf{x}}
\safemath{\bmy}{\mathbf{y}}
\safemath{\bmz}{\mathbf{z}}
\safemath{\bmzero}{\mathbf{0}}
\safemath{\bmone}{\mathbf{1}}

\bmdefine{\biad}{a}
\bmdefine{\bibd}{b}
\bmdefine{\bicd}{c}
\bmdefine{\bidd}{d}
\bmdefine{\bied}{e}
\bmdefine{\bifd}{f}
\bmdefine{\bigd}{g}
\bmdefine{\bihd}{h}
\bmdefine{\biid}{i}
\bmdefine{\bijd}{j}
\bmdefine{\bikd}{k}
\bmdefine{\bild}{l}
\bmdefine{\bimd}{m}
\bmdefine{\bind}{n}
\bmdefine{\biod}{o}
\bmdefine{\bipd}{p}
\bmdefine{\biqd}{q}
\bmdefine{\bird}{r}
\bmdefine{\bisd}{s}
\bmdefine{\bitd}{t}
\bmdefine{\biud}{u}
\bmdefine{\bivd}{v}
\bmdefine{\biwd}{w}
\bmdefine{\bixd}{x}
\bmdefine{\biyd}{y}
\bmdefine{\bizd}{z}

\bmdefine{\bixid}{\xi}
\bmdefine{\bilambdad}{\lambda}
\bmdefine{\bimud}{\mu}
\bmdefine{\bithetad}{\theta}
\bmdefine{\biphid}{\phi}
\bmdefine{\bideltad}{\delta}

\safemath{\bmia}{\biad}
\safemath{\bmib}{\bibd}
\safemath{\bmic}{\bicd}
\safemath{\bmid}{\bidd}
\safemath{\bmie}{\bied}
\safemath{\bmif}{\bifd}
\safemath{\bmig}{\bigd}
\safemath{\bmih}{\bihd}
\safemath{\bmii}{\biid}
\safemath{\bmij}{\bijd}
\safemath{\bmik}{\bikd}
\safemath{\bmil}{\bild}
\safemath{\bmim}{\bimd}
\safemath{\bmin}{\bind}
\safemath{\bmio}{\biod}
\safemath{\bmip}{\bipd}
\safemath{\bmiq}{\biqd}
\safemath{\bmir}{\bird}
\safemath{\bmis}{\bisd}
\safemath{\bmit}{\bitd}
\safemath{\bmiu}{\biud}
\safemath{\bmiv}{\bivd}
\safemath{\bmiw}{\biwd}
\safemath{\bmix}{\bixd}
\safemath{\bmiy}{\biyd}
\safemath{\bmiz}{\bizd}

\safemath{\bmxi}{\bixid}
\safemath{\bmlambda}{\bilambdad}
\safemath{\bmmu}{\bimud}
\safemath{\bmtheta}{\bithetad}
\safemath{\bmphi}{\biphid}
\safemath{\bmdelta}{\bideltad}

\safemath{\bA}{\mathbf{A}}
\safemath{\bB}{\mathbf{B}}
\safemath{\bC}{\mathbf{C}}
\safemath{\bD}{\mathbf{D}}
\safemath{\bE}{\mathbf{E}}
\safemath{\bF}{\mathbf{F}}
\safemath{\bG}{\mathbf{G}}
\safemath{\bH}{\mathbf{H}}
\safemath{\bI}{\mathbf{I}}
\safemath{\bJ}{\mathbf{J}}
\safemath{\bK}{\mathbf{K}}
\safemath{\bL}{\mathbf{L}}
\safemath{\bM}{\mathbf{M}}
\safemath{\bN}{\mathbf{N}}
\safemath{\bO}{\mathbf{O}}
\safemath{\bP}{\mathbf{P}}
\safemath{\bQ}{\mathbf{Q}}
\safemath{\bR}{\mathbf{R}}
\safemath{\bS}{\mathbf{S}}
\safemath{\bT}{\mathbf{T}}
\safemath{\bU}{\mathbf{U}}
\safemath{\bV}{\mathbf{V}}
\safemath{\bW}{\mathbf{W}}
\safemath{\bX}{\mathbf{X}}
\safemath{\bY}{\mathbf{Y}}
\safemath{\bZ}{\mathbf{Z}}

\safemath{\bZero}{\mathbf{0}}
\safemath{\bOne}{\mathbf{1}}
\safemath{\bDelta}{\mathbf{\Delta}}
\safemath{\bLambda}{\mathbf{\UpLambda}}
\safemath{\bPhi}{\mathbf{\Upphi}}
\safemath{\bSigma}{\mathbf{\Upsigma}}
\safemath{\bOmega}{\mathbf{\Upomega}}
\safemath{\bTheta}{\mathbf{\Uptheta}}

\bmdefine{\biAd}{A}
\bmdefine{\biBd}{B}
\bmdefine{\biCd}{C}
\bmdefine{\biDd}{D}
\bmdefine{\biEd}{E}
\bmdefine{\biFd}{F}
\bmdefine{\biGd}{G}
\bmdefine{\biHd}{H}
\bmdefine{\biId}{I}
\bmdefine{\biJd}{J}
\bmdefine{\biKd}{K}
\bmdefine{\biLd}{L}
\bmdefine{\biMd}{M}
\bmdefine{\biOd}{N}
\bmdefine{\biPd}{O}
\bmdefine{\biQd}{P}
\bmdefine{\biRd}{R}
\bmdefine{\biSd}{S}
\bmdefine{\biTd}{T}
\bmdefine{\biUd}{U}
\bmdefine{\biVd}{V}
\bmdefine{\biWd}{W}
\bmdefine{\biXd}{X}
\bmdefine{\biYd}{Y}
\bmdefine{\biZd}{Z}

\bmdefine{\biDelta}{\Delta}
\bmdefine{\biLambda}{\Lambda}
\bmdefine{\biPhi}{\Phi}
\bmdefine{\biSigma}{\Sigma}
\bmdefine{\biOmega}{\Omega}
\bmdefine{\biTheta}{\Theta}

\safemath{\bimA}{\biAd}
\safemath{\bimB}{\biBd}
\safemath{\bimC}{\biCd}
\safemath{\bimD}{\biDd}
\safemath{\bimE}{\biEd}
\safemath{\bimF}{\biFd}
\safemath{\bimG}{\biGd}
\safemath{\bimH}{\biHd}
\safemath{\bimI}{\biId}
\safemath{\bimJ}{\biJd}
\safemath{\bimK}{\biKd}
\safemath{\bimL}{\biLd}
\safemath{\bimM}{\biMd}
\safemath{\bimN}{\biNd}
\safemath{\bimO}{\biOd}
\safemath{\bimP}{\biPd}
\safemath{\bimQ}{\biQd}
\safemath{\bimR}{\biRd}
\safemath{\bimS}{\biSd}
\safemath{\bimT}{\biTd}
\safemath{\bimU}{\biUd}
\safemath{\bimV}{\biVd}
\safemath{\bimW}{\biWd}
\safemath{\bimX}{\biXd}
\safemath{\bimY}{\biYd}
\safemath{\bimZ}{\biZd}

\safemath{\bimDelta}{\biDelta}
\safemath{\bimLambda}{\biLambda}
\safemath{\bimPhi}{\biPhi}
\safemath{\bimSigma}{\biSigma}
\safemath{\bimOmega}{\biOmega}
\safemath{\bimTheta}{\biTheta}

\safemath{\setA}{\mathcal{A}}
\safemath{\setB}{\mathcal{B}}
\safemath{\setC}{\mathcal{C}}
\safemath{\setD}{\mathcal{D}}
\safemath{\setE}{\mathcal{E}}
\safemath{\setF}{\mathcal{F}}
\safemath{\setG}{\mathcal{G}}
\safemath{\setH}{\mathcal{H}}
\safemath{\setI}{\mathcal{I}}
\safemath{\setJ}{\mathcal{J}}
\safemath{\setK}{\mathcal{K}}
\safemath{\setL}{\mathcal{L}}
\safemath{\setM}{\mathcal{M}}
\safemath{\setN}{\mathcal{N}}
\safemath{\setO}{\mathcal{O}}
\safemath{\setP}{\mathcal{P}}
\safemath{\setQ}{\mathcal{Q}}
\safemath{\setR}{\mathcal{R}}
\safemath{\setS}{\mathcal{S}}
\safemath{\setT}{\mathcal{T}}
\safemath{\setU}{\mathcal{U}}
\safemath{\setV}{\mathcal{V}}
\safemath{\setW}{\mathcal{W}}
\safemath{\setX}{\mathcal{X}}
\safemath{\setY}{\mathcal{Y}}
\safemath{\setZ}{\mathcal{Z}}
\safemath{\emptySet}{\varnothing}

\safemath{\colA}{\mathscr{A}}
\safemath{\colB}{\mathscr{B}}
\safemath{\colC}{\mathscr{C}}
\safemath{\colD}{\mathscr{D}}
\safemath{\colE}{\mathscr{E}}
\safemath{\colF}{\mathscr{F}}
\safemath{\colG}{\mathscr{G}}
\safemath{\colH}{\mathscr{H}}
\safemath{\colI}{\mathscr{I}}
\safemath{\colJ}{\mathscr{J}}
\safemath{\colK}{\mathscr{K}}
\safemath{\colL}{\mathscr{L}}
\safemath{\colM}{\mathscr{M}}
\safemath{\colN}{\mathscr{N}}
\safemath{\colO}{\mathscr{O}}
\safemath{\colP}{\mathscr{P}}
\safemath{\colQ}{\mathscr{Q}}
\safemath{\colR}{\mathscr{R}}
\safemath{\colS}{\mathscr{S}}
\safemath{\colT}{\mathscr{T}}
\safemath{\colU}{\mathscr{U}}
\safemath{\colV}{\mathscr{V}}
\safemath{\colW}{\mathscr{W}}
\safemath{\colX}{\mathscr{X}}
\safemath{\colY}{\mathscr{Y}}
\safemath{\colZ}{\mathscr{Z}}

\safemath{\opA}{\mathbb{A}}
\safemath{\opB}{\mathbb{B}}
\safemath{\opC}{\mathbb{C}}
\safemath{\opD}{\mathbb{D}}
\safemath{\opE}{\mathbb{E}}
\safemath{\opF}{\mathbb{F}}
\safemath{\opG}{\mathbb{G}}
\safemath{\opH}{\mathbb{H}}
\safemath{\opI}{\mathbb{I}}
\safemath{\opJ}{\mathbb{J}}
\safemath{\opK}{\mathbb{K}}
\safemath{\opL}{\mathbb{L}}
\safemath{\opM}{\mathbb{M}}
\safemath{\opN}{\mathbb{N}}
\safemath{\opO}{\mathbb{O}}
\safemath{\opP}{\mathbb{P}}
\safemath{\opQ}{\mathbb{Q}}
\safemath{\opR}{\mathbb{R}}
\safemath{\opS}{\mathbb{S}}
\safemath{\opT}{\mathbb{T}}
\safemath{\opU}{\mathbb{U}}
\safemath{\opV}{\mathbb{V}}
\safemath{\opW}{\mathbb{W}}
\safemath{\opX}{\mathbb{X}}
\safemath{\opY}{\mathbb{Y}}
\safemath{\opZ}{\mathbb{Z}}
\safemath{\opZero}{\mathbb{O}}
\safemath{\identityop}{\opI}

\safemath{\veca}{\bma}
\safemath{\vecb}{\bmb}
\safemath{\vecc}{\bmc}
\safemath{\vecd}{\bmd}
\safemath{\vece}{\bme}
\safemath{\vecf}{\bmf}
\safemath{\vecg}{\bmg}
\safemath{\vech}{\bmh}
\safemath{\veci}{\bmi}
\safemath{\vecj}{\bmj}
\safemath{\veck}{\bmk}
\safemath{\vecl}{\bml}
\safemath{\vecm}{\bmm}
\safemath{\vecn}{\bmn}
\safemath{\veco}{\bmo}
\safemath{\vecp}{\bmp}
\safemath{\vecq}{\bmq}
\safemath{\vecr}{\bmr}
\safemath{\vecs}{\bms}
\safemath{\vect}{\bmt}
\safemath{\vecu}{\bmu}
\safemath{\vecv}{\bmv}
\safemath{\vecw}{\bmw}
\safemath{\vecx}{\bmx}
\safemath{\vecy}{\bmy}
\safemath{\vecz}{\bmz}

\safemath{\veczero}{\bmzero}
\safemath{\vecone}{\bmone}
\safemath{\vecxi}{\bmxi}
\safemath{\veclambda}{\bmlambda}
\safemath{\vecmu}{\bmmu}
\safemath{\vectheta}{\bmtheta}
\safemath{\vecphi}{\bmphi}
\safemath{\vecdelta}{\bmdelta}

\safemath{\matA}{\bA}
\safemath{\matB}{\bB}
\safemath{\matC}{\bC}
\safemath{\matD}{\bD}
\safemath{\matE}{\bE}
\safemath{\matF}{\bF}
\safemath{\matG}{\bG}
\safemath{\matH}{\bH}
\safemath{\matI}{\bI}
\safemath{\matJ}{\bJ}
\safemath{\matK}{\bK}
\safemath{\matL}{\bL}
\safemath{\matM}{\bM}
\safemath{\matN}{\bN}
\safemath{\matO}{\bO}
\safemath{\matP}{\bP}
\safemath{\matQ}{\bQ}
\safemath{\matR}{\bR}
\safemath{\matS}{\bS}
\safemath{\matT}{\bT}
\safemath{\matU}{\bU}
\safemath{\matV}{\bV}
\safemath{\matW}{\bW}
\safemath{\matX}{\bX}
\safemath{\matY}{\bY}
\safemath{\matZ}{\bZ}
\safemath{\matzero}{\bmzero}

\safemath{\matDelta}{\bDelta}
\safemath{\matLambda}{\bLambda}
\safemath{\matPhi}{\bPhi}
\safemath{\matSigma}{\bSigma}
\safemath{\matOmega}{\bOmega}
\safemath{\matTheta}{\bTheta}

\safemath{\matidentity}{\matI}
\safemath{\matone}{\matO}

\safemath{\rnda}{A}
\safemath{\rndb}{B}
\safemath{\rndc}{C}
\safemath{\rndd}{D}
\safemath{\rnde}{E}
\safemath{\rndf}{F}
\safemath{\rndg}{G}
\safemath{\rndh}{H}
\safemath{\rndi}{I}
\safemath{\rndj}{J}
\safemath{\rndk}{K}
\safemath{\rndl}{L}
\safemath{\rndm}{M}
\safemath{\rndn}{N}
\safemath{\rndo}{O}
\safemath{\rndp}{P}
\safemath{\rndq}{Q}
\safemath{\rndr}{R}
\safemath{\rnds}{S}
\safemath{\rndt}{T}
\safemath{\rndu}{U}
\safemath{\rndv}{V}
\safemath{\rndw}{W}
\safemath{\rndx}{X}
\safemath{\rndy}{Y}
\safemath{\rndz}{Z}

\safemath{\rveca}{\bimA}
\safemath{\rvecb}{\bimB}
\safemath{\rvecc}{\bimC}
\safemath{\rvecd}{\bimD}
\safemath{\rvece}{\bimE}
\safemath{\rvecf}{\bimF}
\safemath{\rvecg}{\bimG}
\safemath{\rvech}{\bimH}
\safemath{\rveci}{\bimI}
\safemath{\rvecj}{\bimJ}
\safemath{\rveck}{\bimK}
\safemath{\rvecl}{\bimL}
\safemath{\rvecm}{\bimM}
\safemath{\rvecn}{\bimN}
\safemath{\rveco}{\bomO}
\safemath{\rvecp}{\bimP}
\safemath{\rvecq}{\bimQ}
\safemath{\rvecr}{\bimR}
\safemath{\rvecs}{\bimS}
\safemath{\rvect}{\bimT}
\safemath{\rvecu}{\bimU}
\safemath{\rvecv}{\bimV}
\safemath{\rvecw}{\bimW}
\safemath{\rvecx}{\bimX}
\safemath{\rvecy}{\bimY}
\safemath{\rvecz}{\bimZ}

\safemath{\rvecxi}{\bmxi}
\safemath{\rveclambda}{\bmlambda}
\safemath{\rvecmu}{\bmmu}
\safemath{\rvectheta}{\bmtheta}
\safemath{\rvecphi}{\bmphi}

\safemath{\rmatA}{\bimA}
\safemath{\rmatB}{\bimB}
\safemath{\rmatC}{\bimC}
\safemath{\rmatD}{\bimD}
\safemath{\rmatE}{\bimE}
\safemath{\rmatF}{\bimF}
\safemath{\rmatG}{\bimG}
\safemath{\rmatH}{\bimH}
\safemath{\rmatI}{\bimI}
\safemath{\rmatJ}{\bimJ}
\safemath{\rmatK}{\bimK}
\safemath{\rmatL}{\bimL}
\safemath{\rmatM}{\bimM}
\safemath{\rmatN}{\bimN}
\safemath{\rmatO}{\bimO}
\safemath{\rmatP}{\bimP}
\safemath{\rmatQ}{\bimQ}
\safemath{\rmatR}{\bimR}
\safemath{\rmatS}{\bimS}
\safemath{\rmatT}{\bimT}
\safemath{\rmatU}{\bimU}
\safemath{\rmatV}{\bimV}
\safemath{\rmatW}{\bimW}
\safemath{\rmatX}{\bimX}
\safemath{\rmatY}{\bimY}
\safemath{\rmatZ}{\bimZ}

\safemath{\rmatDelta}{\bimDelta}
\safemath{\rmatLambda}{\bimLambda}
\safemath{\rmatPhi}{\bimPhi}
\safemath{\rmatSigma}{\bimSigma}
\safemath{\rmatOmega}{\bimOmega}
\safemath{\rmatTheta}{\bimTheta}

\newenvironment{textbmatrix}{	\setlength{\arraycolsep}{2.5pt}%
								\big[\begin{matrix}}{\end{matrix}\big]%
								\raisebox{0.08ex}{\vphantom{M}}}
								
\def\be{\begin{equation}}
\def\ee{\end{equation}}
\def\een{\nonumber \end{equation}}
\def\mat{\begin{bmatrix}}
\def\emat{\end{bmatrix}}
\def\btm{\begin{textbmatrix}}
\def\etm{\end{textbmatrix}}

\def\ba#1\ea{\begin{align}#1\end{align}}
\def\bas#1\eas{\begin{align*}#1\end{align*}}
\def\bs#1\es{\begin{split}#1\end{split}} 
\def\bg#1\eg{\begin{gather}#1\end{gather}}
\def\bml#1\eml{\begin{multline}#1\end{multline}}
\def\bi#1\ei{\begin{itemize}#1\end{itemize}} 

\newcommand{\lefto}{\mathopen{}\left}

\DeclareMathOperator{\sign}{sgn}			
\DeclareMathOperator*{\argmin}{arg\;min}		
\DeclareMathOperator{\Exop}{\opE}			

\newcommand{\Ex}[2]{\ensuremath{\Exop_{#1}\lefto[#2\right]}} 	

\newcommand{\vecnorm}[1]{\lefto\lVert#1\right\rVert}		

\safemath{\dirac}{\delta}					
\safemath{\krond}{\dirac}					

\safemath{\upto}{\uparrow}
\safemath{\downto}{\downarrow}
\safemath{\iu}{j}							
\safemath{\ev}{\lambda}						
\safemath{\hilseqspace}{l^{2}}				
\newcommand{\banachfunspace}[1]{\setL^{#1}}	
\safemath{\hilfunspace}{\banachfunspace{2}}	

\safemath{\SNR}{\textsf{SNR}} 				
\safemath{\PAR}{\textsf{PAR}} 				
\safemath{\No}{N_0}							
\safemath{\Es}{E_s}							
\safemath{\Eb}{E_b}							
\safemath{\EbNo}{\frac{\Eb}{\No}}
\safemath{\EsNo}{\frac{\Es}{\No}}

\DeclareMathOperator{\CHop}{\ensuremath{\opH}} 
\safemath{\tvir}{\rndh_{\CHop}}				
\safemath{\tvtf}{\rndl_{\CHop}}				
\safemath{\spf}{\rnds_{\CHop}}				
\safemath{\bff}{H_{\CHop}}					

\safemath{\ircf}{r_{h}}						
\safemath{\tftvcf}{r_{s}}					
\safemath{\tfcf}{r_{l}}						
\safemath{\bfcf}{r_{H}}						

\safemath{\tcorr}{c_h}						
\safemath{\scf}{c_{s}}						
\safemath{\tfcorr}{c_{l}}					
\safemath{\fcorr}{c_{H}}						

\safemath{\mi}{I}							
\safemath{\capacity}{C}						

\safemath{\normal}{\mathcal{N}}			
\safemath{\jpg}{\mathcal{CN}}			
\safemath{\mchain}{\leftrightarrow}		

\safemath{\dB}{\,\mathrm{dB}}
\safemath{\dBm}{\,\mathrm{dBm}}
\safemath{\Hz}{\,\mathrm{Hz}}
\safemath{\kHz}{\,\mathrm{kHz}}
\safemath{\MHz}{\,\mathrm{MHz}}
\safemath{\GHz}{\,\mathrm{GHz}}
\safemath{\s}{\,\mathrm{s}}
\safemath{\ms}{\,\mathrm{ms}}
\safemath{\mus}{\,\mathrm{\text{\textmu}s}}
\safemath{\ns}{\,\mathrm{ns}}
\safemath{\ps}{\,\mathrm{ps}}
\safemath{\meter}{\,\mathrm{m}}
\safemath{\mm}{\,\mathrm{mm}}
\safemath{\cm}{\,\mathrm{cm}}
\safemath{\m}{\,\mathrm{m}}
\safemath{\W}{\,\mathrm{W}}
\safemath{\mW}{\, \mathrm{mW}}
\safemath{\J}{\,\mathrm{J}}
\safemath{\K}{\,\mathrm{K}}
\safemath{\bit}{\,\mathrm{bit}}
\safemath{\nat}{\,\mathrm{nat}}

\safemath{\define}{\triangleq}			

\safemath{\equivalent}{\sim}
\safemath{\distas}{\sim}					
\safemath{\sdiff}{\Delta}				

\safemath{\reals}{\mathbb{R}}
\safemath{\positivereals}{\reals_{+}}
\safemath{\integers}{\mathbb{Z}}
\safemath{\posint}{\integers_{+}}
\safemath{\naturals}{\mathbb{N}}
\safemath{\posnaturals}{\naturals_{+}}
\safemath{\complexset}{\mathbb{C}}
\safemath{\rationals}{\mathbb{Q}}

\newcommand*{\fancyrefapplabelprefix}{app}		
\newcommand*{\fancyrefthmlabelprefix}{thm}		
\newcommand*{\fancyreflemlabelprefix}{lem}		
\newcommand*{\fancyrefcorlabelprefix}{cor}		
\newcommand*{\fancyrefdeflabelprefix}{def}		
\newcommand*{\fancyrefproplabelprefix}{prop}	
\newcommand*{\fancyrefobslabelprefix}{obs}		
\newcommand*{\fancyrefalglabelprefix}{alg}		
\newcommand*{\fancyrefasmlabelprefix}{asm}	    
\newcommand*{\fancyreftbllabelprefix}{tbl}	    

\frefformat{vario}{\fancyrefseclabelprefix}{Section~#1}
\frefformat{vario}{\fancyrefthmlabelprefix}{Theorem~#1}
\frefformat{vario}{\fancyreflemlabelprefix}{Lemma~#1}
\frefformat{vario}{\fancyrefcorlabelprefix}{Corollary~#1}
\frefformat{vario}{\fancyrefdeflabelprefix}{Definition~#1}
\frefformat{vario}{\fancyrefobslabelprefix}{Observation~#1}
\frefformat{vario}{\fancyrefasmlabelprefix}{Assumption~#1}
\frefformat{vario}{\fancyreffiglabelprefix}{Figure~#1}
\frefformat{vario}{\fancyrefapplabelprefix}{Appendix~#1} 
\frefformat{vario}{\fancyrefproplabelprefix}{Proposition~#1}
\frefformat{vario}{\fancyrefalglabelprefix}{Algorithm~#1}
\frefformat{vario}{\fancyrefeqlabelprefix}{(#1)}
\frefformat{vario}{\fancyreftbllabelprefix}{Table~#1}

\newtheorem{rem}{Remark}

\safemath{\dictab}{[\,\dicta\,\,\dictb\,]}

\safemath{\ysig}{\bmy}
\safemath{\ysighat}{\hat{\ysig}}
\safemath{\ysigdim}{M}
\safemath{\xsig}{\bmx}
\safemath{\xsigdim}{N}
\safemath{\nx}{n_x}
\safemath{\zsig}{\bmz}
\safemath{\zsigdim}{\ysigdim}
\safemath{\rsig}{\bmr}
\safemath{\Adict}{\bA}
\safemath{\Adicttilde}{\widetilde{\Adict}}
\safemath{\Adictdim}{\outputdim\times\xsigdim}
\safemath{\avec}{\bma}
\safemath{\avectilde}{\tilde{\avec}}
\safemath{\Bdict}{\bB}
\safemath{\Bdicttilde}{\widetilde{\Bdict}}
\safemath{\Cdict}{\bC}
\safemath{\cvec}{\bmc}
\safemath{\Ddict}{\bD}
\safemath{\Ddictdim}{\ysigdim\times\xsigdim}
\safemath{\dvec}{\bmd}
\safemath{\Ddicttilde}{\widetilde{\bD}}
\safemath{\Bonb}{\bB}
\safemath{\bvec}{\bmb}
\safemath{\Bonbdim}{\ysigdim\times\ysigdim}
\safemath{\noise}{\bmn}
\safemath{\noisedim}{\ysigim}
\safemath{\err}{\bme}
\safemath{\errdim}{\ysigdim}
\safemath{\errset}{\setE}
\safemath{\nerr}{n_e}
\safemath{\delop}{\bP_\errset}
\safemath{\delopc}{\bP_{{\errset}^c}}
\safemath{\cplxi}{\imath}
\safemath{\cplxj}{\jmath}

\safemath{\dict}{\matD}
\safemath{\inputdim}{N}		
\safemath{\outputdim}{M}		
\safemath{\sparsity}{S}	
\safemath{\inputdimA}{{N_a}}	
\safemath{\inputdimB}{{N_b}}	
\safemath{\elemA}{{n_a}}	
\safemath{\elemB}{{n_b}}	
\safemath{\resA}{\matR_a}	
\safemath{\resB}{\matR_b}	
\safemath{\subD}{\matS} 
\safemath{\subA}{\matS_a} 
\safemath{\subB}{\matS_b} 
\safemath{\dicta}{\matA} 	
\safemath{\dictb}{\matB} 	
\safemath{\hollowS}{H}
\safemath{\hollowA}{H_a}
\safemath{\hollowB}{H_b}
\safemath{\cross}{Z}
\safemath{\coh}{\mu_d}			
\safemath{\coha}{\mu_a}			
\safemath{\cohb}{\mu_b}			
\safemath{\mubs}{\nu}	
\safemath{\cohm}{\mu_m} 
\safemath{\dictset}{\setD}	
\safemath{\dictsetp}{\dictset(\coh,\coha,\cohb)}	
\safemath{\dictsetgen}{\dictset_\text{gen}}
\safemath{\dictsetgenp}{\dictsetgen(\coh)}
\safemath{\dictsetonb}{\dictset_\text{onb}}
\safemath{\dictsetonbp}{\dictsetonb(\coh)}

\safemath{\leftside}{U}
\safemath{\rightsideA}{R_a}
\safemath{\rightsideB}{R_b}

\safemath{\indexS}{\setI_S} 

\safemath{\na}{n_a}			
\safemath{\nb}{n_b}			
\safemath{\coeffa}{p_i}	
\safemath{\coeffb}{q_j}	
\safemath{\seta}{\setP}		
\safemath{\setb}{\setQ}     
\safemath{\setw}{\setW}	
\safemath{\setz}{\setZ}	
\safemath{\cola}{\veca}		
\safemath{\colb}{\vecb}		
\safemath{\cold}{\vecd}		
\safemath{\inputvec}{\vecx} 	
\safemath{\error}{\vece}	
\safemath{\noiseout}{\vecz} 	
\safemath{\inputvecel}{x}
\safemath{\inputveca}{\vecx_a}
\safemath{\inputvecb}{\vecx_b}
\safemath{\outputvec}{\vecy}	
\safemath{\lambdamin}{\lambda_{\mathrm{min}}}

\safemath{\elltwo}{\ell_2}
\safemath{\ellone}{\ell_1}
\safemath{\ellzero}{\ell_0}
\safemath{\ellinf}{\ell_\infty}
\safemath{\ellinftilde}{\ell_{\widetilde\infty}}
\safemath{\licard}{Z(\coh,\coha,\cohb)}
\safemath{\xsol}{\hat{x}}
\safemath{\xbord}{x_b}		
\safemath{\xstat}{x_s}		
\safemath{\xstatLone}{\tilde{x}_s}
\safemath{\order}{\mathcal{O}} 
\safemath{\scales}{\Theta} 
\safemath{\ones}{\mathbf{1}} 
\safemath{\zeroes}{\mathbf{0}} 
\safemath{\thlone}{\kappa(\coh,\cohb)} 
\safemath{\constoneA}{\delta} 
\safemath{\constoneB}{\epsilon} 
\safemath{\nlarge}{L}				   
\safemath{\sumlarge}{S_\nlarge}
\safemath{\maxlarger}{P_\nlarge}	   
\safemath{\Pzero}{\textrm{P0}}	
\safemath{\Pone}{\textrm{P1}}
\safemath{\vecfir}{\vecw}			 
\safemath{\vecsec}{\vecz}
\safemath{\elvecfir}{w}              
\safemath{\elvecsec}{z}				 
\safemath{\nlargefir}{n}
\safemath{\normout}{\gamma}
\safemath{\auxfun}{h}
\safemath{\supp}{\textrm{supp}}

\safemath{\indexa}{\ell}
\safemath{\indexb}{r}
\safemath{\indexc}{i}
\safemath{\indexd}{j}

\safemath{\project}{P}

\newcommand{\OPP}{{\text{OPP}}}
\newcommand{\BCRp}{{\text{BCR}$^\ast$}}

\newcommand{\tallm}{\overline{\bH}^\Upsilon}
\newcommand{\tallmp}{\widetilde{\bH}^\Upsilon}

\newtheorem{alg}{Algorithm}
\newtheorem{theorem}{Theorem}

\newcommand{\snr}{\varrho} 

\IEEEoverridecommandlockouts 
\allowdisplaybreaks
\usepackage{color}

\graphicspath{{./figs/}}

\begin{document}

\title{1-bit Massive MU-MIMO Precoding in VLSI}
\author{Oscar Casta\~neda, Sven Jacobsson, Giuseppe Durisi,  \\ Mikael Coldrey, Tom Goldstein, and Christoph Studer\thanks{O.~Casta\~neda and C.~Studer are with the School of Electrical and Computer Engineering, Cornell University, Ithaca, NY (e-mail:  \url{oc66@cornell.edu}, \url{studer@cornell.edu}; web: \url{vip.ece.cornell.edu}).}\thanks{S.\ Jacobsson is with Ericsson Research and Chalmers University of Technology, Gothenburg, Sweden (e-mail: \url{sven.jacobsson@ericsson.com}).}\thanks{G.\ Durisi is with Chalmers University of Technology, Gothenburg, Sweden (e-mail: \url{durisi@chalmers.se}).}\thanks{M.\ Coldrey is with Ericsson Research, Gothenburg, Sweden (e-mail: \url{mikael.coldrey@ericsson.com})}\thanks{T. Goldstein is with the Department of Computer Science, University of Maryland, College Park, MD (e-mail: \url{ tomg@cs.umd.edu}).}
\thanks{The C1PO algorithm implemented in this paper builds upon the 1-bit precoding algorithm {presented} at the IEEE International Conference on Acoustics, Speech, and Signal Processing {(ICASSP)} \cite{castaneda17icassp}; in contrast to the algorithm in~\cite{castaneda17icassp}, C1PO directly operates in the complex domain, comes with convergence guarantees, and can be implemented efficiently in VLSI.}\thanks{A MATLAB simulator for the precoders proposed in this paper is available on GitHub: \url{https://github.com/quantizedmassivemimo/1bit_precoding_VLSI}}}

\maketitle


\begin{abstract}
Massive multiuser (MU) multiple-input multiple-output (MIMO) will be a core technology in fifth-generation~(5G) wireless systems as it offers significant improvements in spectral efficiency compared to existing multi-antenna technologies.
The presence of hundreds of antenna elements at the base station~(BS), however, results in excessively high hardware costs and power consumption, and requires high interconnect throughput between the baseband-processing unit and the radio unit.
Massive MU-MIMO that uses low-resolution analog-to-digital and digital-to-analog converters (DACs) has the potential to address all these issues.
In this paper, we focus on downlink precoding for massive MU-MIMO systems with 1-bit DACs at the BS.
The objective is to design precoders that simultaneously mitigate multi-user interference (MUI) and quantization artifacts.
We propose two nonlinear 1-bit precoding algorithms and corresponding very-large scale integration (VLSI) designs. 
Our algorithms rely on biconvex relaxation, which enables the design of efficient 1-bit precoding algorithms that achieve superior error-rate performance compared to that of linear precoding algorithms followed by quantization. 
To showcase the efficacy of our algorithms, we design VLSI architectures that enable efficient 1-bit precoding for massive MU-MIMO systems in which hundreds of antennas serve tens of user equipments. 
We present corresponding field-programmable gate array~(FPGA) reference implementations to demonstrate that 1-bit precoding enables reliable and high-rate downlink data transmission in practical systems.
\end{abstract}





\section{Introduction}
\label{sec:introduction}

\IEEEPARstart{M}{assive} multiuser (MU) multiple-input multiple-output (MIMO) is widely believed to be a core technology in fifth-generation (5G) wireless systems as it enables substantial improvements in spectral efficiency and reliability compared to traditional, small-scale MIMO technology~\cite{rusek14a, larsson14a, lu14a}. 
These advantages are a result of equipping the base station~(BS) with hundreds or thousands of antennas, which enables fine-grained beamforming to serve tens of user equipments~(UEs) in the same time-frequency resource.
However, the large number of antenna elements and radio frequency (RF) chains at the BS results in a significant increase in hardware complexity, system costs, and circuit power consumption. 
Furthermore, massive MU-MIMO requires high interconnect and chip input/output~(I/O) bandwidth between the baseband-processing unit at the BS and the radio units~\cite{li16globalsip,li2017decentralized}.
As a consequence, {a successful deployment of this technology in 5G wireless systems requires novel design approaches that jointly reduce  system costs, power consumption, and interconnect bandwidth without degrading the spectral efficiency and link reliability.}
\subsection{Massive MU-MIMO with 1-bit DACs}
We consider the massive MU-MIMO downlink in which the BS is equipped with 1-bit digital-to-analog converters (DACs) and transmits data to multiple  UEs in the same time-frequency resource.
In traditional multi-antenna BSs, each RF port is connected to a pair of high-resolution DACs (e.g., with 10-bit precision). Scaling such architectures to massive MIMO BSs, with hundreds or thousands of antennas would result in prohibitively high power consumption and system costs. The deployment of 1-bit DACs at the BS would mitigate this problem.
In addition, the use of 1-bit DACs enables one to lower the linearity and noise requirements of the surrounding RF circuitry, which has the potential to additionally reduce the circuit power consumption.
Another benefit of using 1-bit DACs is the fact that lowering their resolution also reduces the {interconnect} bandwidth between the baseband-processing unit and the radio unit{, as only one bit per sample is required by each DAC.} 
This aspect is of practical relevance for deployment scenarios in which these two units are not co-located~\cite{li16globalsip,li2017decentralized}.

The key challenges of 1-bit massive MU-MIMO systems are to maintain high spectral efficiency and reliability. 
{
The work in~\cite{jacobsson16c} demonstrates that the performance degradation caused by 1-bit DACs in the downlink diminishes as the number of BS antennas increases.}
{Furthermore, as} shown in~\cite{jacobsson16c,jacobsson16b,jedda16a,tirkkonen17a,castaneda17icassp}, the use of 1-bit DACs in the downlink enables reliable data transmission if sophisticated precoding algorithms that simultaneously mitigate multi-user interference~(MUI) and quantization artifacts are used. 
While conventional linear precoding methods, such as zero-forcing~(ZF) or minimum mean-{squared} error (MMSE) precoding followed by quantization, require low computational complexity~\cite{mezghani09c,saxena16a,guerreiro16a,usman16a}, more sophisticated, nonlinear methods are necessary to enable reliable communication at high spectral efficiency. Such precoding methods, however, typically require high computational complexity.
{As a consequence, a successful deployment of 1-bit massive MU-MIMO calls for the design of novel and efficient precoding algorithms that can be implemented in hardware and reliably achieve high throughput at low power consumption.}
\subsection{Contributions}
In this paper, we develop novel, computationally efficient precoding algorithms for 1-bit massive MU-MIMO systems and corresponding  very-large scale integrated (VLSI) designs. 
Our main contributions can be summarized as follows: 
\begin{itemize}
\item We use biconvex relaxation (BCR) \cite{shah2016biconvex} to design a nonlinear 1-bit precoding algorithm. Our algorithm, referred to as C1PO (short for biConvex 1-bit PrecOding), enables reliable, high-rate downlink transmission in 1-bit massive MU-MIMO systems for medium-sized antenna arrays. 
\item We propose a  scalable and low-complexity algorithm variant, referred to as C2PO, which enables high-performance 1-bit precoding for massive MU-MIMO systems with hundreds or thousands of antenna elements. 
\item For C1PO and C2PO, which both solve nonconvex problems, we provide analytical convergence guarantees. 
\item We develop two massively parallel VLSI architectures that implement C1PO and C2PO, and achieve high throughputs in a hardware-efficient manner. Our architectures support various BS and UE antenna configurations. 
\item We present reference designs on a Xilinx Virtex-7 field-programmable gate array (FPGA) for various antenna configurations that demonstrate the efficacy of our algorithms and VLSI architectures. 
\item We compare our designs to a baseline precoder that uses maximum ratio {transmission (MRT)} followed by quantization {(MRT-Q)}, a method that achieves high hardware-efficiency at the cost of poor error-rate performance.
\item We study the trade-offs between error-rate performance and hardware efficiency (in terms of throughput per area) for the proposed FPGA designs.
\end{itemize}
Our results demonstrate the practical feasibility of 1-bit precoding in massive MU-MIMO systems, supporting reliable and high-rate downlink data transmission.

\subsection{Relevant Prior Art}

A number of papers have studied the use of low-resolution analog-to-digital converters (ADCs) for the massive MU-MIMO uplink (UEs transmit data to the BS) with particular focus on the 1-bit case; see, e.g.,~\cite{risi14a,jacobsson15a,li16b,mollen16c,studer15a} and the references therein.
All these results have shown that the use of 1-bit ADCs is sufficient for reliable low-rate uplink transmission and  that~$4$ to $6$~bits are sufficient to close the gap to the infinite-precision case in most scenarios.
In contrast to the uplink, the quantized downlink has gained attention only recently. 
{Precoding in the downlink with 1-bit DACs is a more challenging problem as both MUI and quantization artifacts must be mitigated simultaneously.}
The results in~\cite{mezghani09c,saxena16a,guerreiro16a,usman16a} have shown that so-called \emph{linear-quantized} precoders, which perform traditional linear precoding followed by quantization, enable reliable {downlink} transmission for very large BS antenna arrays in the high signal-to-noise ratio (SNR) regime, even for systems that use 1-bit DACs.  
More sophisticated, \emph{nonlinear} precoding algorithms have been proposed only recently in~\cite{jacobsson16c,jacobsson16b,jedda16a,tirkkonen17a,castaneda17icassp} and significantly outperform linear-quantized methods in the {presence} of 1-bit DACs. The computational complexity of these algorithms, however, is typically high, which prevents an efficient implementation in practical systems.
In contrast to these precoding methods, we propose two novel nonlinear precoding algorithms and VLSI designs that achieve high throughput in a hardware-efficient manner.

While a large number of VLSI designs for data detection in the massive MU-MIMO uplink have been proposed in the literature (see, e.g., \cite{WYWDCS2014,wu2016efficient,wu2016high,castaneda2016data} and references therein), only a handful of precoder designs for multi-antenna downlink systems exist \cite{barrenechea2010design,prabhu2014hardware,shepard2013practical,prabhu2017isscc,li16globalsip}. Reference \cite{barrenechea2010design} proposes a VLSI design for vector-perturbation precoding in small-scale MIMO systems with {high-precision} DACs. The papers~\cite{prabhu2014hardware} and~\cite{shepard2013practical} discuss hardware implementations for approximate linear and ZF/{MRT}-based precoding, respectively, for massive MU-MIMO systems with {high-precision} DACs. Unfortunately, both of these publications do not provide detailed FPGA implementation results. 
{Reference \cite{prabhu2017isscc} describes an application specific integrated circuit (ASIC) design of a ZF precoder; reference \cite{li16globalsip} presents a decentralized ZF precoder on a graphics processing unit (GPU) cluster.
Both of these precoders are, however, designed for high-precision DACs and not for 1-bit massive MU-MIMO systems.}
Hence, to the best of our knowledge, the VLSI designs proposed in this paper are the first hardware implementations reported in the open literature that are suitable for {precoding in the 1-bit massive MU-MIMO downlink.}

\subsection{Notation}
Lowercase and uppercase boldface letters designate column vectors and matrices, respectively. 
For a matrix $\bA$, we denote its transpose, Hermitian transpose, complex conjugate, and matrix $\ell_2$-norm by~$\bA^T$, $\bA^H$, $\bA^*$, and $\|\bA\|_{2,2}$, respectively; the entry on the $k$th row and on the $\ell$th column of $\bA$ is $[\matA]_{k,\ell}$. 
The $M\times M$ identity matrix is denoted by  $\bI_M$ and the $M\times N$ all-zeros matrix is denoted by  $\bZero_{M\times N}$.
For a vector $\veca$, the $k$th entry is~$[\veca]_k$ and we use~$\vecnorm{\veca}_2$ to denote the $\ell_2$-norm of the vector $\bma$.
The real and imaginary parts of a complex vector $\veca$ are $\Re\{\veca\}$ and $\Im\{\veca\}$, respectively. 
The signum function~$\text{sgn}(\cdot)$  is defined as~$\text{sgn}(a)=+1$ for~$a\ge0$ and~$\text{sgn}(a)=-1$ for~$a<0$ and is applied entry-wise to vectors.
The multivariate complex-valued circularly-symmetric Gaussian probability density function (PDF) with  covariance matrix~$\bK$ is denoted by~$\setC\setN(\bZero,\bK)$. We use~$\Ex{\bmx}{\cdot}$ to denote expectation with respect to the random vector $\bmx$.

\subsection{Paper Outline}

The rest of this paper is organized as follows. 
In \fref{sec:systemmodel}, we introduce the system model and formulate the precoding problem for systems with 1-bit DACs.
In \fref{sec:biconvexrelaxation}, we propose two new 1-bit precoding algorithms, namely C1PO and C2PO. 
In \fref{sec:architecture1} and \fref{sec:architecture2}, we detail our VLSI architectures for C1PO and C2PO, respectively. 
In \fref{sec:results}, we show numerical simulations, reference FPGA implementation results, and a comparison with an {MRT}-based baseline precoder. 
We conclude the paper in \fref{sec:conclusions}.
All proofs are relegated to Appendices \ref{app:c1po} and~\ref{app:c2po}.


\begin{figure*}[t]
\centering
 \includegraphics[width=0.85\textwidth]{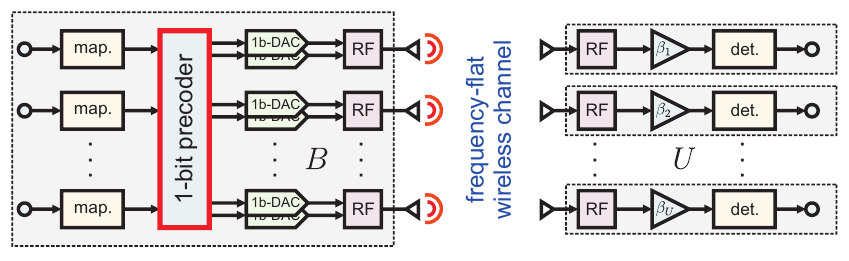}
 \caption{Overview of {an uncoded} massive MU-MIMO downlink system with 1-bit DACs. Left: $B$ antenna massive MU-MIMO BS containing a 1-bit precoder that mitigates multi-user interference and quantization artifacts in the  1-bit DACs; Right: $U$ single-antenna UEs. } 
\label{fig:system_overview}
\end{figure*}
\section{System Model and 1-bit Precoding}
\label{sec:systemmodel}

We start by introducing the downlink system model and then provide the necessary details about optimal precoding in 1-bit massive MU-MIMO systems.
  
\subsection{Downlink System Model}

We focus on the downlink of a single-cell, narrowband massive MU-MIMO system as illustrated in \fref{fig:system_overview}. The system consists of a $B$-antenna BS that serves $U\leq B$ single-antenna\footnote{For simplicity, we focus on single-antenna UEs; the model can easily be expanded to support multi-antenna UEs.} UEs simultaneously and in the same frequency band.
We use the standard input-output relation $\vecy = \matH \vecx + \vecn$ to model the narrowband downlink channel~\cite{rusek14a}.
Here, the vector $\vecy = [y_1,\,\dots,\,y_U]^T$ contains the received signals at all UEs, where $y_u \in \complexset$ is the signal received at the $u$th UE. 
The matrix~$\matH \in \opC^{U \times B}$ represents the downlink channel.
The so-called \emph{precoded vector} is denoted by $\bmx\in\setX^B$, where~$\setX$ represents the transmit alphabet; this set coincides with the set~$\opC$ of complex numbers in the case of infinite-precision DACs. In 1-bit massive MU-MIMO systems, the in-phase and quadrature components are {generated} separately using a pair of 1-bit DACs running at Nyquist rate and hence, the per-antenna quaternary transmit alphabet is~$\setX=\{\pm\ell\pm j\ell\}$ for a given (and fixed) \mbox{$\ell>0$} that determines the transmit power.
The vector~$\bmn\in\complexset^U$ models  i.i.d.\ circularly-symmetric complex Gaussian noise with variance $N_0$ per complex entry, i.e.,~$n_u\sim \jpg(0,N_0)$, for~$u = 1,\,\dots,\,U$. 
In what follows, we assume that the realization of the channel matrix $\bH$ and the noise variance $\No$ are perfectly known at the~BS.\footnote{Knowledge of $\bH$ is typically acquired via training in the uplink in a time-division duplexing system~\cite{rusek14a}. As discussed in \cite{jacobsson16c}, channel estimation errors yield only a small performance loss. Knowledge of the noise variance~$N_0$ at the BS can  be obtained by explicit feedback from the UEs to the BS.}

\subsection{Precoding Basics}

The main purpose of precoding is to transmit the constellation points $s_u \in \setO$ to each UE~$u=1,\ldots,U$, where~$\setO$ is the constellation set (e.g., 16-QAM).
The BS uses the available channel state information (CSI)  to precode the symbol vector $\vecs = [s_1,\,\dots,\,s_U]^T$ into the precoded vector $\vecx \in \setX^B$. Throughout the paper, we assume that the precoded vector~$\vecx$ must satisfy an instantaneous power constraint
$\vecnorm{\vecx}^2_2 = P$; this leads to $\setX = \big\{\pm \ell \pm j \ell\big\}$ with $\ell=\sqrt{P/(2B)}$. 

Coherent transmission of data using multiple BS antennas leads to an {array gain}, which depends on the realization of the fading channel and the precoding method. As in \cite{jacobsson16c,jacobsson16b}, we assume that the $u$th UE is able to rescale its received signals~$y_u$ by a factor\footnote{In contrast to references~\cite{jacobsson16c,jacobsson16b}, which {assumed} real-valued factors $\beta_u$, $u=1,\ldots,U$, we allow these factors to be complex-valued.}  $\beta_u \in \complexset$ in order to compute an estimate~$\hat{s}_u = \beta_u y_u$ for $u=1,\ldots,U$ of the transmitted symbol $s_u\in\setO$.

Since the UEs cannot perform joint processing to recover the transmitted data, precoding must simultaneously reduce MUI and increase signal power at  all UEs~\cite{bjornson14b}. To accomplish these goals, there exist multiple formulations of this optimization problem based on different performance metrics, e.g., sum-rate throughput or {error-rate} (see~\cite{bjornson13a} for a survey).
As in~\cite{jacobsson16c,jacobsson16b}, we will focus exclusively on precoders that minimize the mean-{squared} error (MSE) between the estimated symbol vector~$\hat{\vecs} = \big[\hat{s}_1,\ldots,\hat{s}_U\big]^T = \beta\bmy$ and the transmitted symbol vector $\vecs$ given by
\begin{IEEEeqnarray}{rCl} \label{eq:mse}
\Ex{\bmn}{\| \bms - \hat\bms\|_2^2}
= \vecnorm{\bms - \beta\matH\bmx}_2^2 + |\beta|^2U \No,
\end{IEEEeqnarray}
where we restrict ourselves to the case in which the precoder results in the same \emph{precoding factor}~$\beta$ for all UEs. 
Hence, in the remainder of this paper we shall assume that $\beta_u=\beta$ for $u = 1, \ldots, U$.
{With this assumption, the MSE after precoding will roughly be the same for all UEs, which guarantees a certain degree of fairness among the UEs; see~\cite{jacobsson16c} for more details.}
In \cite{jacobsson16b} it is shown that the UEs are able to accurately estimate the precoding factor~$\beta$ using pilot-based transmission in block-fading scenarios. 

In the infinite-precision case, an MSE-optimal linear precoder multiplies the symbol vector~$\vecs$ with a precoding matrix~$\matP \in \opC^{B \times U}$ so that \eqref{eq:mse} is minimized on average over all possible transmit vectors~$\bms$ subject to the power constraint. 
This problem, which has been studied extensively for the case of infinite-precision DACs~\cite{joham05a, shi07a}, enables the design of low-complexity linear precoding algorithms~\cite{rusek14a}.

\subsection{MSE-Optimal 1-bit Precoding Problem}
In the 1-bit case, linear-quantized precoders  perform first linear precoding and then quantize the result to the finite transmit set $\setX^B$ as
\begin{align*}
\vecx  = \sqrt{\frac{P}{2B}}\big(\sign\lefto(\Re\lefto\{\matP\vecs\right\}\right) + j\sign\lefto(\Im\lefto\{\matP\vecs\right\}\right)\!\big)
\end{align*}
for a given precoding matrix $\bP$.
Linear-quantized precoders can be   analyzed theoretically and typically exhibit low complexity~\cite{jacobsson16c}.
However, as recently  shown in~\cite{jacobsson16c,jacobsson16b,jedda16a,tirkkonen17a,castaneda17icassp}, significant performance improvements can be obtained by using sophisticated nonlinear precoding methods.

One way to design such nonlinear precoders is to solve the following MSE-optimal 1-bit precoding problem~(OPP), which simultaneously finds the optimal precoding vector $\vecx^\text{\OPP}$ and the associated precoding factor~$\beta^\text{\OPP}$:
\begin{align*}
\text{(\OPP)} \qquad 
\underset{\bmx \in \setX^{B}\!,\, \beta \in \complexset}{\text{minimize}} \,\, \vecnorm{\bms - \beta \matH\bmx}^2_2 + |\beta|^2 U \No.
\end{align*}
We emphasize that for a fixed value of~$\beta$, the problem~(\OPP) is a closest vector problem that is known to be NP-hard~\cite{agrell02a,fincke85a,verdu89a}; this implies that there exists no known algorithm to solve it efficiently for large values of $B$. 
In~\cite{jacobsson16c,jacobsson16b}, approximate methods for solving~(\OPP) using convex relaxation have been proposed, such as the squared-infinity norm Douglas-Rachford splitting (SQUID) algorithm. Such relaxation-based methods, however, still require high computational complexity, which prevents their deployment in practical systems.


\section{1-bit Precoding via Biconvex Relaxation }
\label{sec:biconvexrelaxation}

Since the problem (\OPP{}) is of combinatorial nature, a brute-force search for a solution is intractable in massive MU-MIMO systems with hundreds of  BS antennas. We next propose two nonlinear precoding algorithms that yield approximate but accurate solutions at low computational complexity. 

\subsection{Approximating (\OPP{})}
To solve (\OPP{}) efficiently, we use the BCR framework put forward in \cite{shah2016biconvex}, which was initially proposed for solving large semidefinite programs that appear in computer vision. 
In order to use this framework, we first simplify the objective function of (\OPP) by assuming that $\No\to0$, i.e., we assume that the system operates in the high-SNR regime. {Note that we make this assumption solely for the purpose of deriving computationally efficient algorithms; we show in \fref{sec:simresults} that our algorithms also work well in the low-SNR regime.}
As in~\cite[Eq.~3]{castaneda17icassp}, we take a leap of faith with the approximation
\begin{align} \label{eq:dummy}
\underset{\bmx\in\setX^B}{\text{min}} \underset{\beta\in\complexset}{\text{min}}\, \vecnorm{\vecs - \beta \matH\vecx}^2_2 \approx 
\underset{\bmx\in\setX^B}{\text{min}}\underset{\alpha\in\complexset}{\text{min}}\,\vecnorm{\alpha\vecs -  \matH\vecx}^2_2.
\end{align}
This approximation can be justified by noting that if we can find a precoded vector $\bmx\in\setX^B$ for which $\bms=\bH\bmx$, then both problems in \fref{eq:dummy} are indeed equivalent.
These approximations allow us to rewrite  (\OPP{}) as follows:
\begin{align*}
\text{(\OPP$^*$)} \qquad 
\hat{\bmx} = \argmin_{\bmx \in \setX^{B},\, \alpha \in \complexset}  \vecnorm{\alpha{\vecs} -  {\matH}{\vecx}}^2_2.
\end{align*}

We next get rid of the parameter $\alpha$ in $\text{(\OPP$^*$)}$. 
For a fixed~$\bmx$, the optimal parameter   $\hat{\alpha}(\bmx)$ that minimizes the objective function of $\text{(\OPP$^*$)}$ is given by
\begin{align*}
\hat{\alpha}(\bmx) = \argmin_{\alpha\in\complexset} \, \vecnorm{\alpha{\vecs} - {\matH}{\vecx}}^2_2 = \frac{{\bms}^H{\bH}{\bmx}}{\|{\bms}\|^2_2}.
\end{align*}
By inserting $\hat{\alpha}(\bmx)$ into the objective function of $\text{(\OPP$^*$)}$, we obtain 
\begin{align} \label{eq:objectivefunction}
\vecnorm{\hat\alpha(\bmx){\vecs} -  {\matH}{\vecx}}^2_2 = \vecnorm{\bA\vecx}^2_2   
\end{align}
with 
\begin{align} \label{eq:Amatrix}
\bA =  \bQ \bH \quad \text{and} \quad \bQ = \bI_U-\frac{\bms\bms^H}{\|\bms\|^2_2},
\end{align}
where the matrix $\bQ\in\complexset^{U\times U}$ is a projection onto the orthogonal complement of the space spanned by the symbol vector $\bms$.
Using~\fref{eq:objectivefunction}, the problem $\text{(\OPP$^*$)}$ can be simplified to
\begin{align*}
\text{(\OPP$^{**}$)} \qquad 
\hat{\bmx} = \argmin_{\bmx \in \setX^{B}} \,\, \vecnorm{\bA\vecx}^2_2,
\end{align*}
which remains to be a closest vector problem. 
Nevertheless, the specific form of $\text{(\OPP$^{**}$)}$ enables us to use BCR to efficiently compute approximate but accurate solutions.

\subsection{Biconvex Relaxation (BCR)} \label{sec:sub_biconvexrelaxation}
To solve $\text{(\OPP$^{**}$)}$ using BCR, we first introduce a copy $\bmz$ of the vector $\bmx$, and replace $\text{(\OPP$^{**}$)}$ with the approximation 
\begin{align*}
\hat{\bmx} = \argmin_{\bmx \in \setX^{B},\,\bmz\in\complexset^{B}} \,\, \vecnorm{\bA\vecz}^2_2  + \gamma\|\bmz-\bmx\|_2^2,
\end{align*}
where $\gamma>0$ is a (fixed) regularization parameter. 
We next relax the nonconvex alphabet constraint $\bmx \in \setX^{B}$ to its convex envelope given by
\begin{align}\label{eq:convex_env}
\setB^B = \bigg\{ \bmc \in \complexset^{B}  \,\bigg|\, & 
 |\Re\{c_b\}| \leq \sqrt{\frac{P}{2B}} , \notag \\
 &  |\Im\{c_b\}| \leq \sqrt{\frac{P}{2B}},\, b=1,\ldots, B \bigg\} {.}
\end{align}
{This relaxation} allows us to convexify the precoding problem as follows:
\begin{align*}
\hat{\bmx} = \argmin_{\bmx \in \setB^{B},\,\bmz\in\complexset^{B}} \,\, \vecnorm{\bA\vecz}^2_2  + \gamma\|\bmz-\bmx\|_2^2,
\end{align*}
{which enables the design of algorithms that converge quickly}.
Unfortunately, solving this {optimization} problem yields, in general,  the all-zeros vector, i.e., $\bmx=\bZero_{B\times1}$.
One of the key ideas of BCR is to force the solution of this new problem to satisfy the constraints in~\eqref{eq:convex_env} with equality. This can be accomplished by including a nonconvex regularization term in the objective that promotes large values of~$\bmx.$ As suggested in~\cite{shah2016biconvex}, we use a negative $\ell_2$-norm term to obtain the following biconvex relaxation optimization problem:
\begin{align*}
\text{(\BCRp)} \quad \hat{\bmx}^\text{BCR} = \!\!\!\argmin_{\bmx \in \setB^{B},\,\bmz\in\complexset^{B}} \, \vecnorm{\bA\vecz}^2_2  + \gamma\|\bmz-\bmx\|_2^2 - \delta\|\bmx\|_2^2,
\end{align*}
where $\delta>0$ is a (fixed) regularization parameter.  {If} $\delta<\gamma,$ {then} the formulation \text{(\BCRp)} is {\em biconvex} (i.e., the minimization with respect to $\bmx$ is convex when $\bmz$ is fixed, and vice versa). 
Robust parameter choices are $\gamma=\|\bA^H\bA\|_{2,2}$ and $\gamma/\delta=2$; see \cite{shah2016biconvex} for more details.
{In practice, we use numerical simulations to tune the parameters $\gamma$ and $\delta$ to further improve the empirical performance of our algorithms.}

\subsection{C1PO: biConvex 1-bit PrecOding}
We {have} noted above that the \text{(\BCRp)} problem is biconvex, meaning that minimization with respect to  $\bmx$ alone (with $\bmz$ fixed) or $\bmz$ alone (with $\bmx$ fixed) is convex.
Hence, as done in \cite{shah2016biconvex}, we can solve the \text{(\BCRp)} problem approximately  using alternating minimization. 
Since the problem is nonconvex, initialization critically affects the performance of our algorithm.
We initialize our algorithm with the {MRT} precoded vector~$\bmx^{(1)}=\bH^H\bms$, which yields excellent performance in practice and can be computed efficiently. 
Then, we solve for~$\bmz$ while holding~$\bmx$ fixed; afterwards, we solve for~$\bmx$ while holding $\bmz$ fixed. Specifically, we repeat the following procedure{:}
\begin{align*}
\bmz^{(t+1)} &= \argmin_{\bmz \in\complexset^{B}}  \|\bA\bmz\|_2^2 + \gamma\|\bmz-\bmx^{(t)}\|_2^2 \\
\bmx^{(t+1)} & = \argmin_{\bmx\in\setB^B}  \gamma\|\bmz^{(t+1)}-\bmx\|_2^2 -\delta\|\bmx\|_2^2
\end{align*}
for $t=1,2,\ldots,t_\text{max}$, where $t_\text{max}$ is the maximum number of iterations. 
Both steps are convex optimization problems  that can be solved efficiently in closed form. Hence, the above iterative procedure reduces to  the following simple algorithm, which we call C1PO (short for biConvex 1-bit PrecOding). 

\begin{oframed}
\begin{alg}[C1PO]\label{alg:C1PO}  
Set $\bA$ as in \fref{eq:Amatrix}, initialize \mbox{$\bmx^{(1)}=\bH^H\bms$}, and fix the parameters $\delta$ and $\gamma$ so that $0<\delta<\gamma$.
Then, for every iteration $t=1,2,\ldots,t_\text{max},$ compute 
\begin{align}
\bmz^{(t+1)} &= (\bI_B + \gamma^{-1}\bA^H\bA)^{-1} \bmx^{(t)} \label{eq:step1}\\
\bmx^{(t+1)} & = \mathrm{proj} (\bmz^{(t+1)}). \label{eq:step2}
\end{align}
Here, the expansion-reprojection operator $\mathrm{proj}(\cdot)$ is  
\begin{align*}
 \mathrm{proj} ( z) = & \,  \sign( \Re\{z\} )\min\!\left\{\!\frac{\gamma}{\gamma-\delta}|\Re\{z\}|, \sqrt{\frac{P}{2B}}\right\}  \\
& + j\sign( \Im\{z\} )\min\!\left\{\!\frac{\gamma}{\gamma-\delta}|\Im\{z\}|, \sqrt{\frac{P}{2B}}\right\}
\end{align*}
and is applied element-wise to the vector $\bmz^{(t+1)}$. In the last iteration $t_\text{max}$, the output $\bmx^{(t_\text{max}+1)}$ of C1PO is quantized to the quaternary alphabet $\setX = \big\{\pm \ell \pm j \ell\big\}$ with \mbox{$\ell=\sqrt{P/(2B)}$} as follows:
\begin{align} \label{eq:outputquantization}
\hat\vecx  =  \sqrt{\frac{P}{2B}} \bigg( & \sign\lefto(\Re\lefto\{\bmx^{(t_\text{max}+1)}\right\}\right) \notag \\
& +  j\sign\lefto(\Im\lefto\{\bmx^{(t_\text{max}+1)}\right\}\right)\! \bigg). \\[-0.7cm] \nonumber
\end{align}
\end{alg}
\end{oframed}

Because C1PO decreases the objective function \text{(\BCRp)} on every variable update, and the objective is bounded from below,  the objective values corresponding to the iterates $\{\bmx^{(t)},\bmz^{(t)}\}$ form a convergent sequence.  However, by exploiting the biconvex structure of our problem, we can prove the following stronger result; the proof is given in \fref{app:c1po}.
\begin{theorem} \label{thm:c1po}
Any limit point of the sequence $\{\bmx^{(t)},\bmz^{(t)}\}$ generated by C1PO is a stationary point of \text{(\BCRp)}. 
\end{theorem} 

\sloppy

The main computations performed by C1PO in \fref{alg:C1PO} are (i) the $B\times B $ matrix inversion \mbox{$\bG = (\bI_{B} + \gamma^{-1}\bA^H\bA)^{-1}$}, which can be computed once during a preprocessing stage, and (ii) the per-iteration matrix-vector multiplication \mbox{$\bmz^{(t+1)} =\bG\bmx^{(t)} $} in step \fref{eq:step1}; the complexity of the projection in step \fref{eq:step2} is negligible. 
Unfortunately, the complexity of the matrix inversion, evaluated in terms of operations,\footnote{For simplicity, we  count the number of complex-valued multiplications to characterize the operation count.} scales roughly with $B^3$ and the complexity of the per-iteration matrix-vector product with $B^2$.
Both of these tasks are particularly inefficient for massive MU-MIMO systems with a large number of BS antennas. Therefore, we next propose an algorithmic variant that avoids both of these issues and whose complexity scales more favorably with the number of BS antennas. 

\fussy

\subsection{Fast Algorithm for Very-Large Systems: C2PO} \label{sec:C2POalgorithm}
To obtain our alternative algorithm, we start from the BCR formulation in \text{(\BCRp)} but rather than introducing the auxiliary variable $\bmz$, we attempt to {directly} solve the following nonconvex {optimization} problem:\footnote{To simplify notation, we have divided both terms in the objective function by a factor of two; this scaling does not affect the result.}
\begin{align}
\hat{\bmx} = \argmin_{\bmx \in \setB^{B}} \,\, \frac{1}{2} \vecnorm{\bA\vecx}^2_2    - \frac{\delta}{2}\|\bmx\|_2^2. \label{eq:directform}
\end{align}
We use forward-backward splitting (FBS)~\cite{GSB14,BT09,goldstein2010high}, a computationally efficient method to solve large convex problems. Since the problem in \fref{eq:directform} is nonconvex, FBS is not guaranteed to converge to the optimal solution. Nevertheless, as shown in \fref{sec:results}, the proposed algorithm performs well in practice.

FBS is an efficient iterative procedure to solve convex optimization problems of the form 
\begin{align*}
\hat\bmx=\argmin_\bmx f(\bmx)+g(\bmx),
\end{align*}
where the function $f$ is smooth and convex, and the function $g$ is convex but not necessarily smooth or bounded. FBS consists of the following iteration~\cite{BT09,GSB14}:
\begin{align*}
\bmz^{(t+1)} & = \bmx^{(t)} - \tau^{(t)} \nabla f(\bmx^{(t)}) \\
\bmx^{(t+1)} & = \text{prox}_g\!\left( \bmz^{(t+1)}  ; \tau^{(t)}  \right)
\end{align*}
for $t=1,2,\ldots,t_\text{max}$ or until convergence. Here, $\nabla f(\bmx)$ is the gradient of the smooth function $f$, and the so-called proximal operator for the function $g$ is defined as follows~\cite{parikh2014proximal}:
\begin{align*}
\text{prox}_g \!\left( \bmz  ; \tau  \right) = \argmin_{\bmx} \left\{ \tau g(\bmx)+\frac{1}{2}\|\bmx-\bmz\|_2^2 \right\}\!.
\end{align*}
The sequence $\{\tau^{(t)}>0\}$ contains suitably chosen step-size parameters.  For the problem \eqref{eq:directform}, we show below that FBS monotonically decreases the objective \eqref{eq:directform} for any constant step size that satisfies $\tau^{(t)}=\tau<\|\bA^H\bA\|^{-1}_{2,2}.$

In order to approximately solve \fref{eq:directform} using FBS, we set 
\begin{align}
f(\bmx)=\frac{1}{2}\vecnorm{\bA\vecx}^2_2 \,\text{ and }\,  g(\bmx) =  \chi\!\left(\bmx \in \setB^{B}\right)- \frac{\delta}{2}\|\bmx\|_2^2 , \label{eq:directsplit}
\end{align}
where $\chi$ is a {\em characteristic function} that is zero if the condition $\bmx \in \setB^{B}$ is met and infinity otherwise. For this choice of $f$ and~$g$, the gradient is given by  $\nabla f(\bmx)=\bA^H\bA\vecx$ and the proximal operator is given by the expansion-reprojection operation 
\begin{align}
& \mathrm{prox}_g ( z) =   \sign( \Re\{z\} )\min\!\left\{\!\frac{1}{1-\tau\delta}|\Re\{z\}|, \sqrt{\frac{P}{2B}}\right\} \nonumber \\
& \quad + j\sign( \Im\{z\} )\min\!\left\{\!\frac{1}{1-\tau\delta}|\Im\{z\}|, \sqrt{\frac{P}{2B}}\right\}\!, \label{eq:proxg}
\end{align}
which is valid for $\tau\delta <1$ and applied element-wise to vectors. 
By using FBS with the above-mentioned ingredients, we obtain the following simple algorithm, which we call C2PO. 

\begin{oframed}
\begin{alg}[C2PO] \label{alg:C2PO} 
Set $\bA$ as in \fref{eq:Amatrix}. Initialize \mbox{$\bmx^{(1)}=\bH^H\bms$} and fix the parameters $\delta$ and $\tau$ so that $\tau\delta < 1$.
Then, for every iteration $t=1,2,\ldots,t_\text{max}$ compute: 
\begin{align}
\bmz^{(t+1)} &= \bmx^{(t)} - \tau \bA^H\bA \bmx^{(t)}  \label{eq:faststep1} \\
\bmx^{(t+1)} & = \mathrm{prox}_g ( {\bmz^{(t+1)}};\tau). \label{eq:faststep2} 
\end{align}
Here, the $\mathrm{prox}_g$ operator is given in \fref{eq:proxg} and is applied element-wise to the vector $\bmz^{(t+1)}$. In the last iteration $t_\text{max}$, the output $\bmx^{(t_\text{max}+1)}$ of C2PO is quantized to the quaternary alphabet $\setX$ as in \fref{eq:outputquantization}. 
\end{alg}
\end{oframed}

The following result shows that C2PO is well behaved, provided that the step size is chosen appropriately; the proof  is given in \fref{app:c2po}.

\begin{theorem} \label{thm:c2po}
Suppose the step size used in C2PO satisfies  $\tau < \|\bA^H\bA\|^{-1}_{2,2},$ and $\tau\delta <1.$  Then, C2PO decreases the objective \eqref{eq:directform} monotonically, and any limit point of the iterates $\{\bmx^{(t)}\}$ is a stationary point.
\end{theorem} 

The most complex operation of C2PO (\fref{alg:C2PO}) is {the} matrix-vector multiplication in step~\fref{eq:faststep1}. 
In contrast to C1PO (\fref{alg:C1PO}), however, this step requires a minimal amount of preprocessing  and can be computed efficiently, especially for large BS antenna arrays. To see this, we rewrite~$\bA^H\bA$, where $\bA$ was given in~\fref{eq:Amatrix}, as follows:
\begin{align*}
\bA^H\bA = \bH^H\bH - \frac{\bH^H\bms\bms^H\bH}{\|\bms\|^2_2} = \bH^H\bH - \bmv\bmv^H = {\tallm} \overline{\bH}.
\end{align*}
Here, $\bmv=\bH^H\bms/\|\bms\|_2$ is a normalized version of the {MRT} vector; {the augmented matrices $\overline{\bH}=[\bH ; \bmv^H ]$ and $\tallm=[\bH^H , -\bmv ]$ are of dimension $(U+1)\times B$ and $B\times (U+1)$, respectively.} With these definitions, we can now simplify step~\fref{eq:faststep1} to
\begin{align}
\bmz^{(t+1)} = \bmx^{(t)} - \tau {\tallm}\overline{\bH} \bmx^{(t)},
\label{eq:simplefaststep1}
\end{align}
where we first compute $\bmw = \overline{\bH} (\tau\bmx^{(t)})$, then $\bmw'={\tallm}\bmw$, and finally $\bmz^{(t+1)} = \bmx^{(t)} - \bmw'$. 
As a result, we conclude that each iteration of C2PO  requires only two matrix-vector products with a cost of roughly $2B(U+1)$ operations (in contrast to $B^2$ operations for C1PO). In addition, the preprocessing stage of this algorithm only needs to compute the normalized {MRT} vector~$\bmv$, which requires roughly $BU$ operations (in contrast to~$B^3$ operations for C1PO). 
Hence, the complexity of C2PO can be significantly lower than that of C1PO, especially since the antenna configurations of typical massive MU-MIMO systems satisfy~$U\ll B$. As we will show in \fref{sec:results}, the hardware efficiency of C2PO is superior to that of C1PO for large BS antenna arrays and the error-rate performance is comparable.

\subsection{Alternative Derivation of C2PO}
It is interesting to note that there is a strong connection between \fref{alg:C1PO} and \fref{alg:C2PO}. In fact, one can obtain C2PO directly from C1PO using the following well-known series expansion. 
Let $\|\bA^H\bA\|_{2,2}<\gamma$. Then, we have the following Neumann series  expansion~\cite{GV96}:
\begin{align*}
 (\bI_B + \gamma^{-1}\bA^H\bA)^{-1} = \sum_{n=1}^\infty (-\gamma^{-1}\bA^H\bA)^n.
\end{align*}
As suggested in~\cite{WYWDCS2014}, we can approximate the inverse by truncating the series to the two first terms:
\begin{align*}
 (\bI_B + \gamma^{-1}\bA^H\bA)^{-1} \approx \bI_B - \gamma^{-1}\bA^H\bA.
\end{align*}
By using this approximation in step \fref{eq:step1} of \fref{alg:C1PO}, we immediately obtain \fref{alg:C2PO} after setting $\gamma^{-1}=\tau$.  Note that the Neumann series expansion is only convergent for $\|\bA^H\bA\|_{2,2}<\gamma,$ which corresponds to the step size restriction $\tau < \|\bA{^H}\bA\|^{-1}_{2,2}$; this is exactly the same step size requirement as in \fref{thm:c2po}.


\section{VLSI Design for C1PO}
\label{sec:architecture1}

We now present a high-throughput VLSI architecture for C1PO as in \fref{alg:C1PO}. We then discuss the key optimizations performed in our FPGA implementation.
\setlength{\textfloatsep}{10pt}
\subsection{Architecture Overview}
The proposed VLSI architecture that implements C1PO as detailed in \fref{alg:C1PO} is shown in \fref{fig:C1POarch}.
Our architecture consists of a linear array of $B$ identical \emph{processing elements (PEs)} that share a common control unit.
The PEs essentially compute the complex-valued matrix-vector product in \eqref{eq:step1}, {using a variant of Cannon's algorithm \cite{CannonThesis},} followed by the projection operation in \eqref{eq:step2}. 
Each PE $b=1,2,\ldots,B$ consists of three main building blocks: (i) a $\bmg_b$-memory, (ii) a complex-valued multiply-accumulate (MAC) unit, and (iii) a projection unit.
For the $b$th PE, the $\bmg_b$-memory stores the $b$th row of the matrix \mbox{$\bG = (\bI_{B} + \gamma^{-1}\bA^H\bA)^{-1}$}, which we assume was computed during a {separate} preprocessing stage. 
{
As mentioned in \fref{sec:sub_biconvexrelaxation}, simulations are used to tune the parameter~$\gamma$ in order to improve the error-rate performance; the optimal value of~$\gamma$ depends on the antenna configuration.}
The complex-valued MAC unit is used by each PE to sequentially compute an entry of the output vector~$\bmz^{(t+1)}$ on line \eqref{eq:step1}, while the entries of the vector~$\bmx^{(t)}$ are exchanged between the PEs in a cyclic fashion; this is done to avoid an architecture with a centralized~$\bmx^{(t)}$ memory that would suffer from a high fan-out because the memory's output has to be distributed to all the PEs.
The projection unit  implements the expansion-reprojection operator $\mathrm{proj}(\cdot)$ on line \eqref{eq:step2} in a hardware-friendly manner.
The outputs of the projection unit are also used to generate the quaternary outputs of the 1-bit precoder; to this end, each PE simply takes the sign bits of the complex-valued output vector~$\bmx^{(t+1)}$.

\subsection{Architecture Operation} \label{sec:arch1operation}
\begin{figure}[tp]
\centering
\vspace{-0.2cm}
\includegraphics[width=0.8\columnwidth]{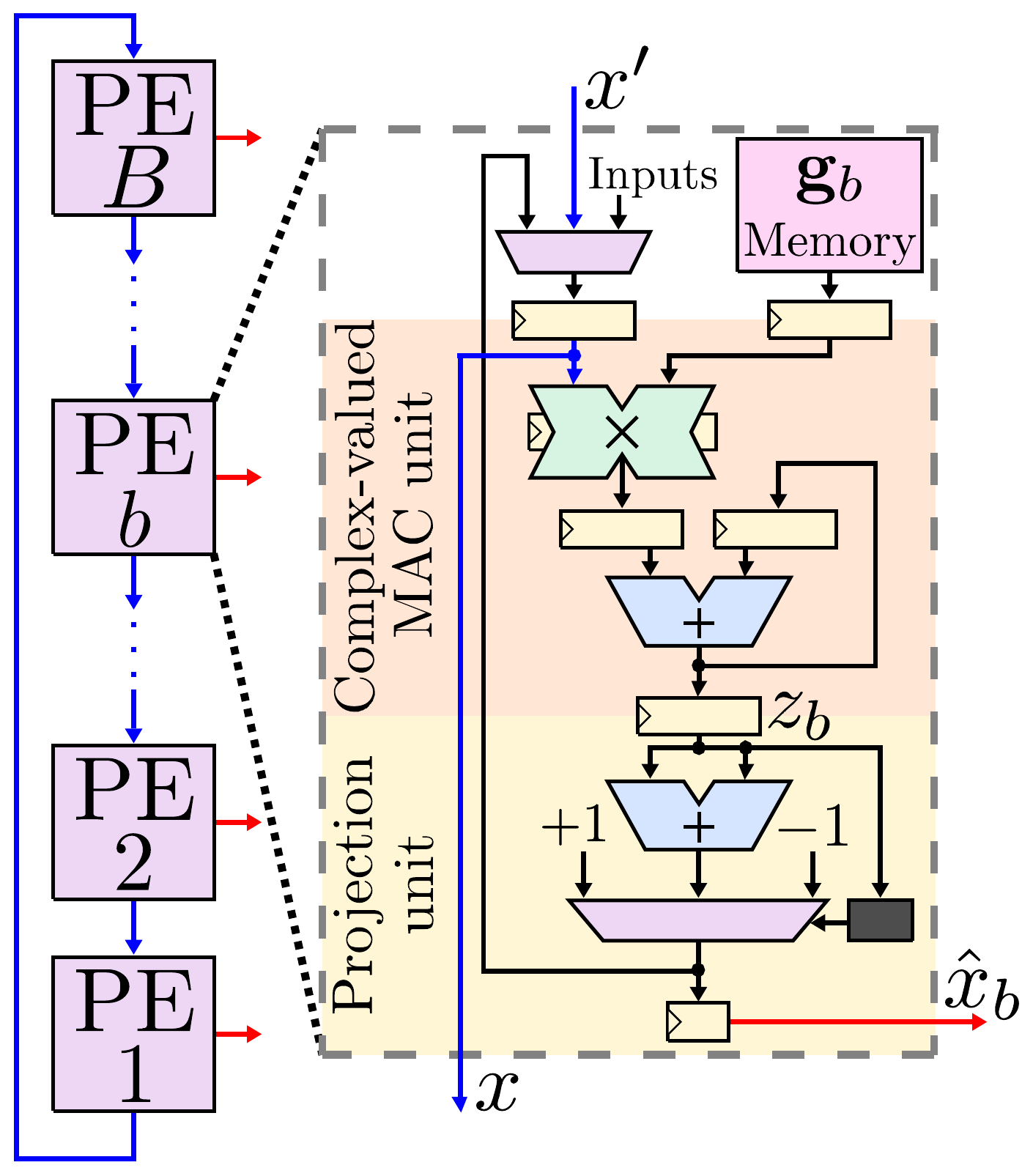}
\vspace{-0.2cm}
\caption{High-level block diagram of the  VLSI architecture for C1PO. We use a linear array of $B$ processing elements~(PEs) that enables us to achieve high throughput at low hardware complexity.}
\label{fig:C1POarch}
\end{figure}
We now detail the (rather technical) operation of the C1PO architecture illustrated in \fref{fig:C1POarch}.
In the first iteration (i.e., at $t=1$), each PE $b$ is initialized with the $b$th entry of the vector $\bmx^{(1)}$.
Furthermore, the entries of the~$\bmg_b$-memory are stored so that the first memory address corresponds to $[\bG]_{b,b}$, the second address to $[\bG]_{b,b+1}$, and so forth (addresses wrap around).

In the first clock cycle, each PE $b$ computes $[\bG]_{b,b}[\bmx^{(t)}]_b$ and the result is stored in the accumulator.
As shown on the left side of \fref{fig:C1POarch}, in the same clock cycle the $b$th PE passes the value $[\bmx^{(t)}]_b$ to PE $(b-1)$, while it receives the value $[\bmx^{(t)}]_{b+1}$ from PE $(b+1)$; PE $1$ passes its value to PE $B$. 
In the second clock cycle, since the exchange operation made the element $[\bmx^{(t)}]_{b+1}$ available at PE $b$, each PE computes $[\bG]_{b,b+1} \cdot [\bmx^{(t)}]_{b+1}$ and uses the accumulator to add it to the result of the previous cycle. 
Once again, in the same clock cycle, the $b$th PE passes the $\bmx^{(t)}$ entry that is currently being multiplied on its MAC unit to PE $(b-1)$; PE $1$ passes its value to PE $B$.
Consequently, in the third clock cycle, the $b$th PE will use the values $[\bG]_{b,b+2}$ and $[\bmx^{(t)}]_{b+2}$ to continue performing MAC operations.
By repeating this procedure $B$ times, each entry of the vector $[\bmx^{(t)}]$ circulates through all the PEs exactly once, enabling each PE $b=1,2,\ldots,B$ to compute $[\bmz^{(t+1)}]_b$. Thus, the matrix-vector multiplication on line \eqref{eq:step1} is completed.
Since the complex-valued MAC unit contains three pipeline stages, two clock cycles are required to flush the pipeline. Hence, the matrix-vector operation requires a latency of $B+2$ clock cycles.
After $B+2$ clock cycles, the vector $\bmz^{(t+1)}$ is available at the outputs of the MAC units. 

In the subsequent clock cycle, each PE projects their respective entry of the vector $\bmz^{(t+1)}$. 
{According to our simulation results,} the choice $\gamma/\delta=5$, which implies that $\gamma/(\gamma-\delta)=1.25$, works well for {all} the considered antenna configurations. 
Furthermore, to reduce the hardware complexity, we assume $P=2B$ so that the clipping threshold of the expansion-reprojection operator $\mathrm{proj}(\cdot)$ is $1$. 
As a result, the $\mathrm{proj}(\cdot)$ operator in \eqref{eq:step2} is implemented by applying the following operations independently to the real and imaginary parts of $[\bmz^{(t+1)}]_b$:
We multiply the real (or imaginary) part of $[\bmz^{(t+1)}]_b$ by $1.25$; this is accomplished by adding the $[\bmz^{(t+1)}]_b$ value with a $2\times$ right-shifted version of itself.
At the same time, the real (or imaginary) part of $[\bmz^{(t+1)}]_b$ is compared to $-0.8$ and $+0.8$.
If the real (or imaginary) part of $[\bmz^{(t+1)}]_b$ is between these two numbers, then the projection unit outputs $1.25 \cdot [\bmz^{(t+1)}]_b$.
If it is smaller than $-0.8$, then the projection unit generates $-1$; if it is larger than $+0.8$, it generates $+1$.
The result from this projection is stored as the next iterate $[\bmx^{(t+1)}]_b$ in the input register of the complex-valued MAC unit, which completes one C1PO iteration.
Since the projection requires an additional clock cycle, one C1PO iteration is completed in exactly $B+3$ clock cycles. 

\subsection{FPGA Implementation Details}
To minimize the FPGA implementation complexity of C1PO, we exclusively use fixed-point arithmetic; see \fref{sec:simresults} for the fixed-point error-rate performance of C1PO. 
To represent the entries of the vector~$\bmx^{(t)}$, we use $12$-bit {signed} fixed-point values with $5$ fraction bits. 
The entries of the $\bG$ matrix are represented using $10$ bits with $9$ fraction bits, and we use FPGA look-up tables (LUTs) as a distributed RAM to store these values.
The complex-valued MAC unit uses $18$ bits with $11$ fraction bits; the projection unit uses $15$ bits with $8$ fraction bits.
In our C1PO design, all adders and multipliers do not saturate, but wrap around; number resizing uses truncation.
All complex-valued multipliers consist of four real-valued multipliers and two adders; we use the built-in DSP48 units for these operations. 


\section{VLSI Design for C2PO}
\label{sec:architecture2}

We now present a high-throughput VLSI architecture for C2PO as in \fref{alg:C2PO}. We then discuss the key optimizations used in our FPGA implementation.

\subsection{Architecture Overview}
\begin{figure*}[tp]
\centering
\vspace{-0.2cm}
\includegraphics[width=0.87\textwidth]{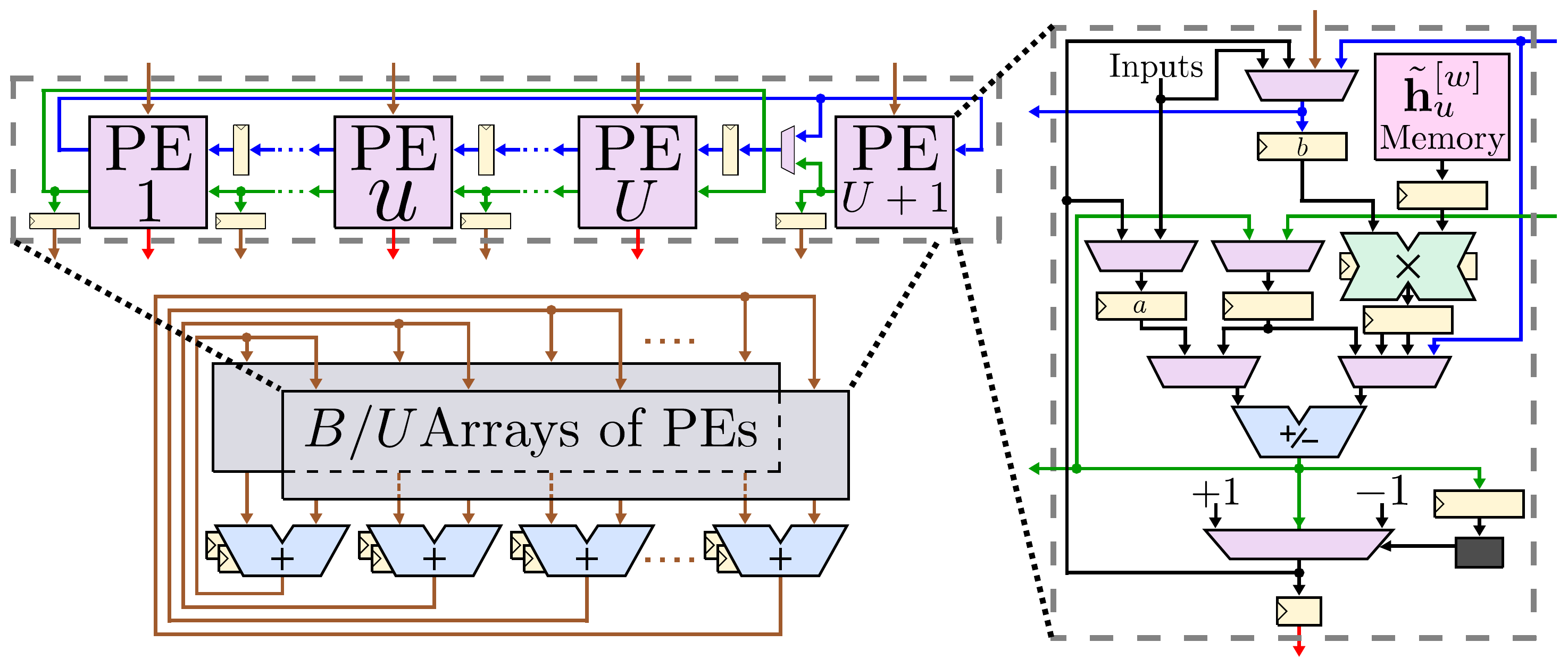}
\vspace{-0.2cm}
\caption{High-level block diagram of the  VLSI architecture for C2PO. We use $B/U$ linear arrays, each consisting of $U+1$ processing elements~(PEs), which enable us to achieve high throughput at low hardware complexity.}
\label{fig:C2POarch}
\end{figure*}
The proposed VLSI architecture that implements C2PO (\fref{alg:C2PO}) is shown in \fref{fig:C2POarch}.
In what follows, we assume that~$B$ is a multiple of $U$ and $B\gg U$. Our architecture consists of $B/U$ linear arrays; each array consists of $U+1$  PEs and a control unit.
The architecture divides the operation in \eqref{eq:simplefaststep1} into two separate matrix-vector products: (i) $\bmw=\overline{\bH}(\tau \bmx^{(t)})$ and (ii) $\bmw'={\tallm} \bmw$; see also the discussion at the end of \fref{sec:C2POalgorithm}. {We assume that $\overline{\bH}$ was computed in a separate preprocessing stage.}
Note that for the first matrix-vector product, the matrix $\overline{\bH}$ has more columns ($B$) than rows ($U+1$); for the second matrix-vector product, the matrix ${\tallm}$ has more rows than columns.
Therefore, we will refer to the first matrix-vector product as the \emph{wide product}, while the second one will be identified as the \emph{tall product}.
The  final subtraction required to compute $\bmz^{(t+1)}=\bmx^{(t)}-\bmw'$ in~\fref{eq:simplefaststep1} is incorporated into the tall-product operation; see \fref{sec:arch2operation} for more details.

To perform both the wide and tall products in a single computation unit, the matrix $\overline{\bH}$ is divided into $B/U$ sub-matrices $\widetilde{\bH}_w \in \complexset^{(U+1)\times U}$, $w=1,2,\ldots,B/U$, so that $\overline{\bH}=\big[\widetilde{\bH}_1,\widetilde{\bH}_2,\ldots,\widetilde{\bH}_{B/U}\big]$.
{In the same way, the matrix $\tallm$ is divided into $B/U$ sub-matrices $\tallmp_w \in \complexset^{U\times (U+1)}$, where $w=1,2,\ldots,B/U$, and $\tallm=\big[\tallmp_1;\tallmp_2;\ldots;\tallmp_{B/U}\big]$.
Note that $\widetilde{\bH}_w^H$ and $\tallmp_w$ are the same matrices, except for a sign flip of the last column.}
Analogously to the matrices case, the vector $\bmx^{(t)}$ is divided into $B/U$ sub-vectors $\tilde{\bmx}^{(t)}_w \in \complexset^U$, $w=1,2,\ldots,B/U$, so that $\bmx^{(t)} = \big[\tilde{\bmx}^{(t)}_1;\tilde{\bmx}^{(t)}_2;\ldots;\tilde{\bmx}^{(t)}_{B/U}\big]$.
We next outline the architectural principle of the wide and tall products. 

{\em (i) Wide Product:}
Each linear array takes one sub-matrix~$\widetilde{\bH}_w$ and the associated sub-vector $\tilde{\bmx}^{(t)}_w$ as its inputs and computes their product in a sequential, column-by-column manner. This operation is analogous to that of the C1PO architecture (cf.~\fref{sec:arch1operation}) and within each linear array, the entries of the scaled sub-vector~$\tau\tilde{\bmx}^{(t)}_w$ are cyclically exchanged among the PEs. 
The resulting vectors $\bmw_w=\widetilde{\bH}_w (\tau \tilde{\bmx}^{(t)}_w)$ are then added  to obtain 
\begin{align*}
\bmw = \overline{\bH}(\tau \bmx^{(t)}) = \sum_{w=1}^{B/U} \widetilde{\bH}_w (\tau \tilde{\bmx}^{(t)}_w),
\end{align*}
which completes the wide product.
Each entry $[\bmw]_u$ of the resulting vector $\bmw$ is then stored in PE $u$ of all linear arrays.

{\em (ii) Tall Product:}
With the $\bmw$ vector available in all the linear arrays, each array now computes $U$ entries of $\bmz^{(t+1)}$ by implementing $\bmz^{(t+1)}_w=\tilde{\bmx}^{(t)}_w-{\tallmp_w} \bmw$. Here, however, we use a sequential procedure in which the accumulated results are exchanged between PEs of the same array. {This procedure is---once again---a variant of Cannon's algorithm \cite{CannonThesis}}; see \fref{sec:arch2operation} for a detailed explanation.
As a result, each linear array can project its computed $\bmz^{(t+1)}_w$ entries to generate the next iterate~$\tilde{\bmx}^{(t+1)}_w$, which are then used by the same linear array to proceed with the next iteration.
The sign bits of the new vector $\bmx^{(t+1)}$ correspond to the outputs of the C2PO architecture.

As in the C1PO architecture, each PE \mbox{$u=1,2,\ldots,U+1$} is formed by three main units.
The first unit is an $\tilde{\bmh}^{[w]}_u$-memory, which stores the $u$th row of the $\widetilde{\bH}_w$ sub-matrix;
{$\tallmp_w$ can be derived directly from $\widetilde{\bH}_w$.}
The second unit is a complex-valued MAC unit, which supports (i) multiplications of $a \times b$ and $a\times b^*$, (ii) accumulation by addition or subtraction, and (iii) initialization of the accumulator with a non-zero value.
The third unit is the projection unit, which is equivalent to the one of C1PO, although it is merged with the accumulator of the MAC unit.

\subsection{Architecture Operation} \label{sec:arch2operation}
We now provide the (rather technical) operation details of the C2PO architecture illustrated in \fref{fig:C2POarch}.
Without loss of generality, we focus our description on the $w$th linear array of PEs, which operates on the $\widetilde{\bH}_w$ sub-matrix and the~$\tilde{\bmx}^{(t)}_w$ sub-vector.
In the first iteration (i.e., at $t=1$), the entry $[\tilde{\bmx}^{(1)}_w]_u$ and its scaled version $\tau [\tilde{\bmx}^{(1)}_w]_u$ are stored in PE $u=1,2,\ldots,U$ in two different registers: The value  $[\tilde{\bmx}^{(1)}_w]_u$ is stored in the register labeled with ``$a$'' in \fref{fig:C2POarch}, which will later be used to initialize the accumulator in the complex-valued MAC unit; the value $\tau [\tilde{\bmx}^{(1)}_w]_u$ is stored at the input register of the MAC unit labeled with ``$b$.'' 
We restrict the stepsize $\tau$ to be {of the form} $2^{-\alpha}$ for some fixed $\alpha\in\mathbb{N}^+$, which enables us to acquire~$\tau [\tilde{\bmx}^{(1)}_w]_u$ from a simple arithmetic right-shifted version of $[\tilde{\bmx}^{(1)}_w]_u${; we used numerical simulations to optimize the error-rate performance by selecting an optimal value for $\tau$}.
The $(U+1)$th PE stores the same value $[\tilde{\bmx}^{(1)}_w]_1$ as that in PE $1$.
Similar to the C1PO architecture, the entries of the~$\tilde{\bmh}^{[w]}_u$-memory are stored so that  the first memory address contains $[\widetilde{\bH}_w]_{u,u}$, the second address $[\widetilde{\bH}_w]_{u,u+1}$, and so forth (addresses wrap around). 
For the $(U+1)$th PE, the first address of the $\tilde{\bmh}^{[w]}_{U+1}$-memory contains $[\widetilde{\bH}_w]_{U+1,1}$, the second address contains $[\widetilde{\bH}_w]_{U+1,2}$, etc.

{\em (i) Wide Product:}
In the first clock cycle, each PE $u=1,2,\ldots,U$ computes $[\widetilde{\bH}_w]_{u,u}\cdot [\tau \tilde{\bmx}^{(t)}_w]_u$ and stores the result in the accumulator.
The $(U+1)$th PE computes $[\widetilde{\bH}_w]_{U+1,1} \cdot [\tau \tilde{\bmx}^{(t)}_w]_1$.
As shown in the upper left side of \fref{fig:C2POarch}, in the same clock cycle, the $u$th PE passes the value $[\tau \tilde{\bmx}^{(t)}_w]_u$ to PE~$(u-1)$, while it receives the value $[\tau \tilde{\bmx}^{(t)}_w]_{u+1}$ from PE $(u+1)$; PE~$1$ passes its value to PE $U$, while PE $(U+1)$ does not pass anything.
In the second clock cycle, since the cyclic exchange operation made the entry $[\tau \tilde{\bmx}^{(t)}_w]_{u+1}$ available at PE $u$, each PE computes $[\widetilde{\bH}_w]_{u,u+1} \cdot [\tau \tilde{\bmx}^{(t)}_w]_{u+1}$ and uses the accumulator to add it to the result of the previous cycle. The $(U+1)$th PE uses the same value $\tau \tilde{\bmx}^{(t)}_w$ as PE 1; hence, it can compute $[\tau \tilde{\bmx}^{(t)}_w]_2 \cdot [\widetilde{\bH}_w]_{U+1,2}$.
Again, in the same clock cycle, the $u$th PE passes the $\tau \tilde{\bmx}^{(t)}_w$ entry that is currently being multiplied on its MAC unit to PE $(u-1)$; PE $1$ passes its value to PE $B$, while PE $(U+1)$ does not pass anything.
Consequently, in the third clock cycle, the $u$th PE will use the values $[\widetilde{\bH}_w]_{u,u+2}$ and $[\tau \tilde{\bmx}^{(t)}_w]_{u+2}$ to continue performing MAC operations. During this third cycle, the $(U+1)$th PE will calculate the product $[\widetilde{\bH}_w]_{U+1,3} \cdot [\tau \tilde{\bmx}^{(t)}_w]_3$.  
By repeating this procedure $U$ times, each entry of the sub-vector $(\tau \tilde{\bmx}^{(t)}_w)$ cycles through all the PEs exactly once, enabling the $w$th linear array of PEs to compute $\widetilde{\bH}_{w} (\tau \tilde{\bmx}^{(t)}_{w})$.
Since the complex-valued MAC unit contains three pipeline stages, two clock cycles are required to flush the pipeline. Hence, the previous matrix-vector operation has a latency of $U+2$ clock cycles.
To complete the wide product, the vectors $\widetilde{\bH}_{w} (\tau \tilde{\bmx}^{(t)}_{w})$  must be added.
We use a binary adder tree with $\log_2(B/U)$ pipeline stages.
Hence, the  vector $\bmw$ is computed after $U+\log_2(B/U)+2$ clock cycles. 
The $u$th PE in each linear array stores the entry $[\bmw]_u$ in the MAC unit's input registered labeled with ``$b$'' in \fref{fig:C2POarch}.

{\em (ii) Tall Product:}
In the next clock cycle, the computation of the tall-product starts.
During the first clock cycle of the tall product computation, the PE $u=1,2,\ldots,U$ has available $[\bmw]_u$, as well as $[\widetilde{\bH}_{w}]_{u,u}$, the first entry in its memory. 
The PE can then compute $[\widetilde{\bH}_{w}]^*_{u,u} \cdot [\bmw]_u = [\widetilde{\bH}^H_{w}]_{u,u} \cdot [\bmw]_u {= [\tallmp_{w}]_{u,u} \cdot [\bmw]_u}$.
Using the accumulator, this product is then subtracted from $[\tilde{\bmx}^{(t)}_w]_u$, which was stored during the initialization phase in the register labeled with ``$a$'' in \fref{fig:C2POarch}.
During the same clock cycle, the $u$th PE sends its accumulated result to the $(u-1)$th PE; PE $1$ sends its accumulated result to PE $U$.
Also, in the same clock cycle, the $(U+1)$th PE multiplies the conjugate of the first entry of its memory with its $\bmw$ entry. In words, the product $[\widetilde{\bH}_w]^*_{U+1,1} \cdot [\bmw]_{U+1} = [\widetilde{\bH}_{w}^H]_{1,U+1} \cdot [\bmw]_{U+1} = {-[\tallmp_{w}]_{1,U+1} \cdot [\bmw]_{U+1}}$ is computed.
The result is sent to the $U$th PE. In the following clock cycles, this result will cycle through the linear array using the same wires and registers that were previously used to transfer the $\tau \tilde{\bmx}^{(t)}_w$ entries.
In the second clock cycle, the $(u-1)$th PE multiplies the value $[\bmw]_{u-1}$ with $[\widetilde{\bH}_w]^*_{u-1,u}={[\tallmp_w]_{u,u-1}}$. The product is then subtracted from the accumulated value received from the $u$th PE during the previous cycle, and the new accumulated value is passed to the $(u-2)$th PE.
In the same clock cycle, the $(U+1)$th PE multiplies the value $[\bmw]_{U+1}$ with $[\widetilde{\bH}_w]^*_{U+1,2}={-[\tallmp_w]_{2,U+1}}$ and sends the result to the $U$th PE, so it can cycle through the linear array. Furthermore, PE $U$ passes the ${-[\tallmp_{w}]_{1,U+1}} \cdot [\bmw]_{U+1}$ (previously received from the $(U+1)$th PE) to PE $(U-1)$.
In the third clock cycle, the $(u-2)$th PE calculates ${[\tallmp_w]_{u,u-2}} \cdot  [\bmw]_{u-2}$, subtracts it from the accumulated result received on the second cycle from the $(u-1)$th PE and passes the result to the $(u-3)$th PE.
In the same clock cycle, the $(U+1)$th PE computes ${-[\tallmp_w]_{3,U+1}} \cdot  [\bmw]_{U+1}$ and sends it to PE $U$. Meanwhile, ${-[\tallmp_{w}]_{2,U+1}} \cdot  [\bmw]_{U+1}$ is passed from PE $U$ to PE $(U-1)$ and ${-[\tallmp_{w}]_{1,U+1}} \cdot [\bmw]_{U+1}$ is passed from PE $(U-1)$ to PE $(U-2)$.
After repeating this procedure for $U$ clock cycles, each PE $u=1,2,\ldots,U$ will contain  the accumulated result for the $u$th entry of $\tilde{\bmx}^{(t)}_w-{\tallmp_w} \bmw$, received from the $(u+1)$th PE during the previous cycle.
However, this accumulated result is missing the product ${-[\tallmp_w]}_{u,U+1} \cdot [\bmw]_{U+1}$, which was computed and sent by the $(U+1)$th PE.
Nonetheless, in the $(U+1)$th cycle of the tall product procedure, the $u$th PE receives the missing ${-[\tallmp_w]}_{u,U+1} \cdot [\bmw]_{U+1}$ value from the $(u+1)$th PE.
{The received product is accumulated with the remaining data by addition instead of subtraction.}
Thus, the $\bmz^{(t+1)}=\bmx^{(t)}-{\tallm} \bmw$ entries are calculated after $U+1$ cycles.
Since the complex MAC unit is used again, two additional clock cycles are required to flush its pipeline. Hence, $U+3$ cycles are used for the tall product. 

Finally, in the subsequent clock cycle after  the tall product is completed, the projection operator is applied in a similar fashion as for the C1PO architecture. 
{As in the C1PO case, our simulation results show that the choice of $\tau\lambda=0.2$  (which implies that $1/(1-\tau\lambda)=1.25$) works well for all the considered antenna configurations.}
{Therefore, the only difference between the projection units of C1PO and C2PO is that, in the C2PO architecture,} the accumulator of the MAC unit is used to multiply the real and imaginary parts of each $\bmz^{(t+1)}$ entry with $1.25$, by adding each $\bmz^{(t+1)}$ entry with a $2\times$ right-shifted version of itself.
The result from this projection is stored as the next iterate $[\tilde{\bmx}^{(t+1)}_w]_u$ in the two initialization locations previously mentioned, completing one C2PO iteration.
Since the projection operation requires an additional clock cycle, a full C2PO iteration is completed in exactly $2U+\log_2(B/U)+6$ clock cycles. 

\subsection{FPGA Implementation Details}
As for the C1PO FPGA implementation, we exclusively use fixed-point arithmetic for the C2PO FPGA design; see \fref{sec:simresults} for the fixed-point error-rate performance of C2PO. 
To represent the entries of the vector $\bmx^{(t)}$, we use $12$-bit {signed} fixed-point values with $5$ fraction bits. For the scaled $\tau \bmx^{(t)}$ values, we use $12$-bit {signed} fixed-point values with $11$ fraction bits.
The entries of the $\overline{\bH}$ matrix consist of $10$\,bits with $8$ fraction bits, and we use FPGA look-up tables (LUTs) as a distributed RAM to store these values.
The complex-valued MAC unit uses $18$ bits with $15$ fraction bits when doing the wide product and $11$ fraction bits when doing the tall product; the projection unit uses $18$ bits with $11$ fraction bits. The adder tree uses $21$ bits with $15$ fraction bits.
Identical to the C1PO implementation, all adders and multipliers do not saturate, but wrap around; number resizing uses truncation.
All complex-valued multipliers are built with four real-valued multipliers and two adders; we use DSP48 units for these operations.


\section{Results}
\label{sec:results}

We now provide error-rate performance results for massive MU-MIMO systems and show reference FPGA implementation results for C1PO and C2PO.

\subsection{Simulation Results} \label{sec:simresults}
\begin{figure*}[tp]
\centering
\subfigure[$B=32$, $U=16$, and BPSK.]{\includegraphics[width=.95\columnwidth]{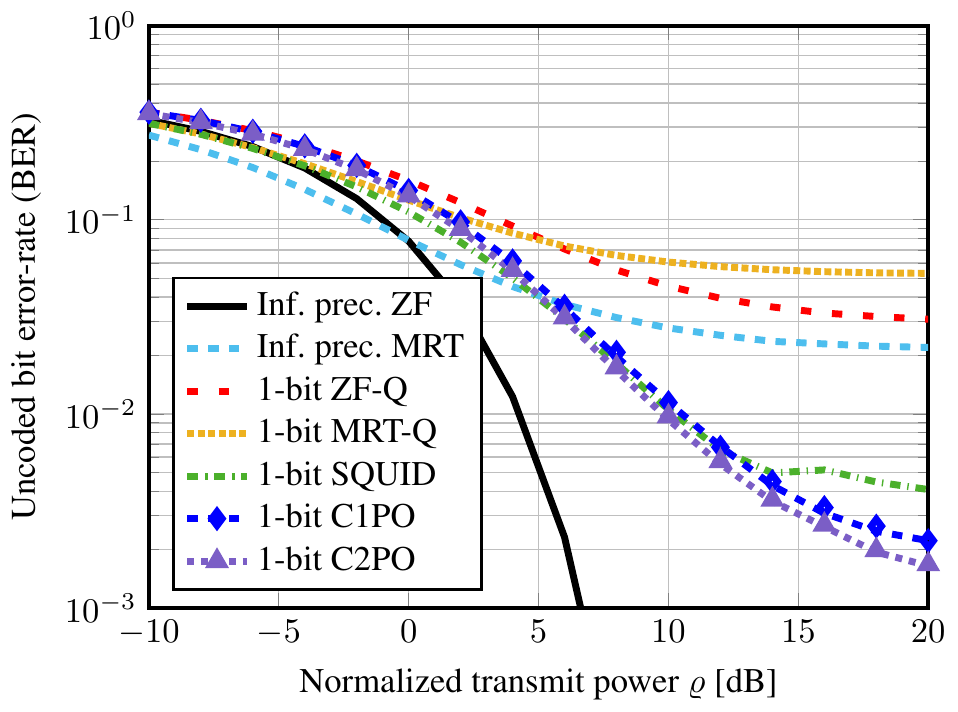}\label{fig:BER_32x16_BPSK}}
\quad
\subfigure[$B=256$, $U=16$, and 16-QAM.]{\includegraphics[width=.95\columnwidth]{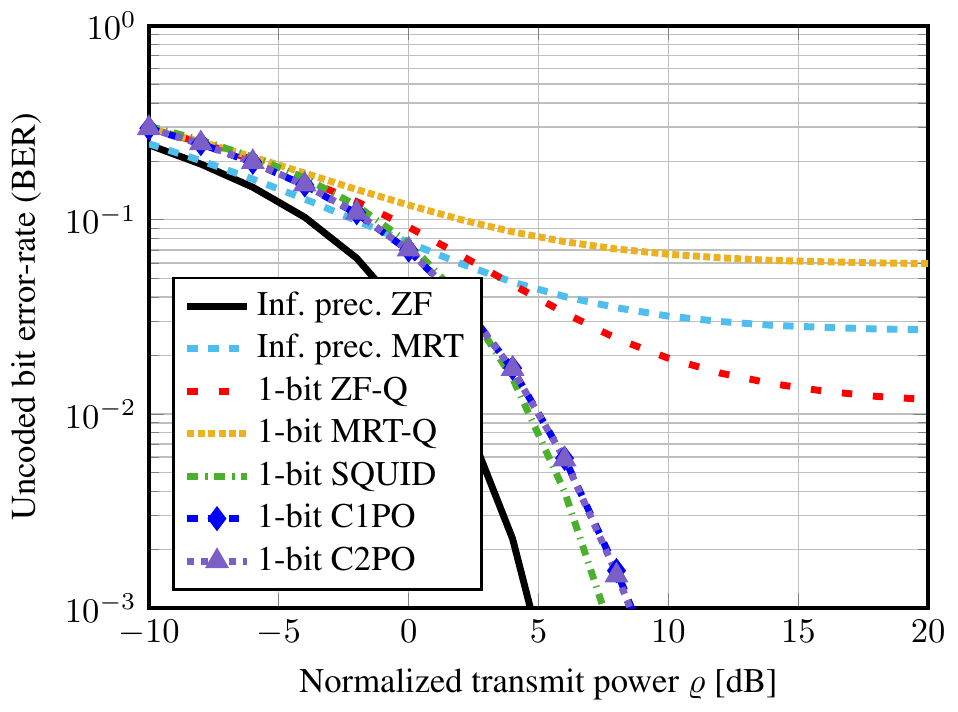}\label{fig:BER_256x16_16QAM}}
\caption{Uncoded bit {error-rate} (BER) for various 1-bit precoders as a function of the normalized transmit power $\snr$ and for different antenna configurations and modulation schemes. C1PO and C2PO achieve similar performance to SQUID~\cite{jacobsson16c}  and significantly outperform linear-quantized precoders, such as quantized zero-forcing (ZF-Q) and {MRT (MRT-Q)}. The performance of ZF {and {MRT}} precoding with infinite-precision DACs {are} included as {references}.}\label{fig:BER_ALL}
\end{figure*}
\fref{fig:BER_ALL} shows uncoded bit {error-rate} (BER) curves versus the normalized transmit power $\snr=P/\No$ for massive MU-MIMO dowlink systems with $U=16$ UEs for various precoding algorithms. 
In \fref{fig:BER_32x16_BPSK} we consider the case of $B=32$ BS antennas with BPSK, whereas in \fref{fig:BER_256x16_16QAM} we consider the case of $B=128$ BS antennas with 16-QAM. For both systems, we use Gray mapping, generate i.i.d.\ Rayleigh fading channel matrices, and average the BER over 10\,000 Monte--Carlo trials.
We compare ZF followed by quantization (ZF-Q), {MRT} followed by quantization {(MRT-Q)}, the nonlinear SQUID algorithm proposed in \cite{jacobsson16c}, as well as C1PO and C2PO, for systems with 1-bit DACs. {As a baseline, we also include ZF and {MRT} with infinite-precision DACs (denoted by ``Inf.~prec.~ZF'' and ``Inf.~prec.~{MRT}'', respectively).}
SQUID runs $t_\text{max}=50$ iterations; C1PO and C2PO both run $t_\text{max}=24$ iterations. 
For all algorithms, the curves represent MATLAB floating-point performance; for C1PO and C2PO, the markers correspond to fixed-point performance of our hardware designs. Clearly, the fixed-point implementation loss of our hardware designs is negligible, i.e., less than 0.15\,dB {normalized transmit power $\snr$} at $1$\% uncoded BER for both considered scenarios. 

For the $16\times32$ system (we use the notation $U\times B$ to refer to a downlink scenario with $U$ users and $B$ BS antennas) with BPSK, \fref{fig:BER_32x16_BPSK} shows that all nonlinear precoders significantly outperform the linear-quantized precoders (ZF-Q and {MRT-Q}), which exhibit a high error floor. Furthermore, we see that C1PO and C2PO achieve similar performance as that of SQUID. At {low values of normalized transmit power $\snr$}, SQUID is marginally better, whereas C2PO achieves the best performance at high {values of $\snr$}, closely followed by C1PO and SQUID. {It can also be seen that, at high values of $\snr$, 1-bit nonlinear precoders significantly outperform the error-rate performance of {MRT} with infinite-precision DACs.}

For the $16\times128$ system with 16-QAM, \fref{fig:BER_256x16_16QAM} shows a similar trend, i.e., non-linear precoders significantly outperform linear-quantized precoders. SQUID outperforms C1PO and C2PO (which perform equally well) by about 0.5\,dB {normalized transmit power $\snr$} at 1\% BER. However, we note that the complexity (in terms of operation counts) of SQUID is more than 2$\times$ higher than that of C1PO and C2PO, and also involves the sorting of $B$ dimensional vectors which is difficult to implement efficiently in VLSI.
We also observe that non-linear precoders enable reliable transmission of higher-order modulation schemes (such as 16-QAM), which is not possible with linear-quantized methods---{the error-rate performance of nonlinear 1-bit precoders for higher-order modulation schemes is studied in more detail in \cite{jacobsson16b}.}
We also see that non-linear precoders do not exhibit an error floor in the considered BER range, which is in contrast to the linear-quantized ones. We note that a detailed theoretical analysis of the error-rate performance of non-linear 1-bit precoders is an open research problem.

\begin{rem}
Our results are limited to a narrowband downlink channel, in which we assume that the BS has perfect knowledge of the channel matrix~$\bH$ and the noise variance $N_0$.
We also assume that all the UEs have approximately the same large-scale fading gains, and we restricted ourselves to a single precoding factor~$\beta$ for all UEs. 
Furthermore, we have ignored real-world hardware impairments and synchronization aspects.
Hence, the provided simulation results are not necessarily representative for other, more realistic system scenarios.
To enable interested readers to perform their own simulations with different system
parameters, we made our MATLAB simulation framework available for download from GitHub: \url{https://github.com/quantizedmassivemimo/1bit_precoding_VLSI}.
\end{rem}

\subsection{FPGA Implementation Results} \label{sec:fpgaresults}
\begin{table*}[tp]
\centering
\renewcommand{\arraystretch}{1.1}
\begin{minipage}[c]{2\columnwidth}
    \centering
    \caption{Implementation results for C1PO and C2PO {for MU-MIMO systems with $U=16$ UEs} on a Xilinx Virtex-7 XC7VX690T~FPGA}
       \label{tbl:implresultsCxPO}
  \begin{tabular}{@{}lccccccccc@{}}
  \toprule
  Algorithm & \multicolumn{4}{c}{C1PO} && \multicolumn{4}{c}{C2PO} \\
  \cmidrule{2-5}  \cmidrule{7-10}
  {BS antennas $B$} & $32$ & $64$ & $128$ & $256$  && $32$ & $64$ & $128$ & $256$\\
  \midrule
  {Slices} & 2\,700 & 5\,187 & 10\,324 & 21\,951 && 3\,375 & 6\,519 & 12\,690 & 24\,748 \\ 
  {LUTs} & 6\,671 & 13\,305 & 30\,979 & 71\,817 && 10\,817 & 21\,920 & 43\,710 & 85\,323 \\
  -- {LUTs as logic} & 6\,031 & 12\,025 & 25\,939 & 51\,897 && 10\,069 & 20\,424 & 40\,718 & 79\,339 \\
  -- {LUTs as memory} & 640 & 1\,280 & 5\,040 & 19\,920 && 748 & 1\,496 & 2\,992 & 5\,984 \\
  {Flipflops} & 6\,830 & 13\,624 & 26\,683 & 52\,175 && 5\,677 & 12\,461 & 26\,083 & 53\,409 \\
  {DSP48 units} & 128 & 256 & 512 & 1\,024 && 136 & 272 & 544 & 1\,088 \\
   \midrule   
  Max.~clock frequency [MHz] & 285 & 264 & 244 & 205 && 222 & 206 & 208 & 193 \\
  {Min.~latency\footnote{The {minimum latency is} measured for one algorithm iteration.} [clock cycles]} & 35 & 67 & 131 & 259 && 39 & 40 & 41 & 42 \\
  {Max.~throughput\footnote{{The throughput corresponds to the total number of symbols precoded per unit of time. In this case, the maximum throughput is equal to $(U f)/ d $, where $f$ is the maximum clock frequency and $d$ the minimum latency.}}\,\,  [Msymbols/s]} & 130 &  63 &  30 & 13 && 91 & 82 &  81 & 74 \\
  {Power consumption\footnote{Statistical power estimation at maximum clock frequency and 1.0\,V supply voltage.} [W]} & 1.13 & 1.97 & 3.43 & 5.74 && 1.04 & 1.70 & 3.17 & 5.80 \\
  \midrule
    {Max.~throughput/LUTs}  & {19\,529} & {4\,733} & {962} & {177} && {8\,413} & {3\,756} & {1\,853} & {862}   \tabularnewline
  \bottomrule
  \end{tabular}
  \end{minipage}
  \end{table*}

To demonstrate the efficacy of C1PO and C2PO, we implemented several FPGA designs for different antenna configurations, namely for $32$, $64$, $128$, and $256$ BS antennas; all designs support downlink transmission to $16$ UEs for modulation schemes ranging from BPSK to 16-QAM. 
The FPGA designs were developed on register transfer level (RTL) using Verilog, implemented using Xilinx Vivado Design Suite, and optimized for a Xilinx Virtex-7 XC7VX690T FPGA.
\fref{tbl:implresultsCxPO} shows reference FPGA implementation results for C1PO and C2PO. 

We see that the logic area (in terms of slices, logic LUTs, flipflops, and DSP48 units) for all designs increases roughly \emph{linearly} with the number of BS antennas; this is a result of using a linear array of PEs. The only exception is the memory requirements of C1PO (in terms of memory LUTs), which scales roughly \emph{quadratically} in the number of BS antennas; this is a result of having to store the entire $B\times B$ matrix~$\bG$ in contrast to storing only the augmented $(U+1)\times B$ matrix~$\overline \bH$ for C2PO. 
We also see that the logic area for C1PO is 20\% to 50\% smaller than that of C2PO for all array sizes; the memory area of C1PO, however, is significantly larger for $128$ and $256$ BS antennas.
This is because the architecture for C1PO is slightly simpler than that of C2PO, but the memory requirements of C1PO scale quadratically in $B$ whereas the memory requirements of C2PO only scale linearly in $B$. 

The maximum clock frequency for C1PO is slightly higher than that of C2PO, which is due to the slightly simpler architecture of C1PO. As expected, the maximum clock frequency slowly decreases with $B$, since the FPGA routing overhead increases with $B$. In fact, after implementing our designs, the critical paths are typically in interconnect networks. Before mapping our designs to the FPGA, however, the critical path for the C1PO designs is in the real-valued multipliers that form part of the complex multiplier, while for the C2PO designs it is in the adders that form part of the complex multiplier.  
The latency of one C1PO iteration is significantly larger than that of C2PO for $64$, $128$, and $256$ antennas. This results in significantly higher throughput of C2PO for these BS antenna array sizes. In summary, C2PO is more efficient in terms of throughput per area for large BS antenna array sizes (e.g, 128 BS antennas or more), whereas C1PO is more efficient for small array sizes.

We finally note that the implementation results in \fref{tbl:implresultsCxPO} ignore the preprocessing complexity in order to compare the complexity of the precoding stage alone. For C1PO, preprocessing requires a $B\times B$ matrix inversion, which is computationally demanding, exhibits stringent data dependencies, and requires high numerical precision, especially for large BS antenna arrays~\cite{WYWDCS2014}. In stark contrast, preprocessing for C2PO only requires the computation of the scaled {MRT} output, which requires a multiplication of a $B\times U$ matrix by a $U$-dimensional vector. As a result, we consider C2PO to be the preferred 1-bit precoding method for most practical BS antenna array sizes.

\subsection{Comparison with {MRT}-based Precoding}
\label{sec:MRCbaselinedesign}
\begin{table}[tp]
\renewcommand{\arraystretch}{1.1}
\begin{minipage}[c]{1\columnwidth}
\vspace{-0.1cm}
    \centering
    \caption{Implementation results for a {MRT-Q}-based precoder for MU-MIMO systems with $U=16$ UEs on~a~Xilinx Virtex-7 XC7VX690T~FPGA}
       \label{tbl:implresultsMRC}
  \begin{tabular}{@{}lcccc@{}}
  \toprule
  BS antennas $B$ & $32$ & $64$ & $128$ & $256$\\
  \midrule
  {Slices} & 2\,543 & 5\,097 & 9\,444 & 17\,630 \\
  {LUTs} & 7\,842 & 15\,617 & 32\,476 & 64\,446 \\
  -- {LUTs as logic} & 7\,010 & 13\,953 & 29\,148 & 57\,790 \\
  -- {LUTs as memory} & 832 & 1\,664 & 3\,328 & 6\,656 \\
  {Flipflops} & 5\,711 & 11\,419 & 21\,902 & 42\,764 \\
  \midrule
  Clock freq. [MHz] & 412 & 410 & 388 & 359 \\
  {Latency [cycles]} & 18 & 18 & 18 & 18 \\
  {TP\footnote{The throughput is calculated as $(U f)/ d $, where $f$ is the clock frequency and $d$ the latency.} [Msymbols/s]} & 366 & 365 & 345 & 319 \\
  {Power\footnote{Statistical power estimation at max.\ clock freq. and 1.0\,V supply voltage.} [W]} & 0.79 & 1.25 & 1.84 & 3.16 \\
  \midrule
  {Throughput/LUTs}  & {46\,665} & {23\,356} & {10\,621} & {4\,945}   \tabularnewline  
  \bottomrule
  \end{tabular}
  \end{minipage}
  \end{table}
While the papers~{\cite{shepard2013practical,prabhu2014hardware,prabhu2017isscc,li16globalsip}} propose hardware designs for precoding in massive MU-MIMO systems with high-precision DACs, neither of them provide detailed FPGA implementation results. 
Reference~\cite{shepard2013practical} describes an FPGA-based testbed that uses {MRT} and ZF-based precoding but does not report area and clock frequency results; {\cite{prabhu2014hardware} and \cite{prabhu2017isscc} only provide} ASIC implementation results{, and \cite{li16globalsip} reports implementation results on a GPU cluster. Furthermore, all of these implementations were designed for high-precision DACs. Consequently, to enable a fair}  comparison of conventional precoders with C1PO and C2PO, we developed a baseline design that implements {MRT} followed by quantization ({MRT-Q}). 

Our {MRT-Q} baseline design is essentially a stripped-down and heavily optimized version of C1PO with only the necessary circuitry to implement {MRT-Q}. More specifically, our architecture corresponds to $B/U$ linear arrays, each one with~$U$ PEs and a control unit. The arrays and PEs are organized as in \fref{fig:C1POarch}, with the exception that the projection unit is removed from the PEs. In addition, no multipliers are required as \mbox{{MRT-Q}} computes~$\bH^H\bms$ with $\bms\in\setO^U,$ and hence all multiplications are with constants (given by the constellation set $\setO$) and can be implemented with adders and shifters. 

The FPGA implementation results for the {MRT-Q} baseline designs are reported in \fref{tbl:implresultsMRC}. Note that these designs do not require any DSP48 units as the multiplication with constants are carried out with conventional logic. In comparison to the 1-bit precoder designs reported in \fref{tbl:implresultsCxPO}, we see that \mbox{{MRT-Q}}-based precoding is roughly $5\times$ to $6\times$ more efficient than C2PO and up to $30\times$ more efficient than C1PO (in terms of throughput/LUTs). This efficiency advantage comes at a significant loss in terms of error-rate performance (cf.~\fref{fig:BER_ALL}). We note, however, that for massive MU-MIMO systems with significantly more BS antennas than UEs (e.g., more than $8\times$), {MRT-Q} is a viable low-complexity alternative---a well-known fact in the massive MU-MIMO literature~\cite{rusek14a, larsson14a, lu14a}.

\subsection{Performance--Complexity Trade-Offs}
\begin{figure}[tp]
\centering
\includegraphics[width=0.95\columnwidth]{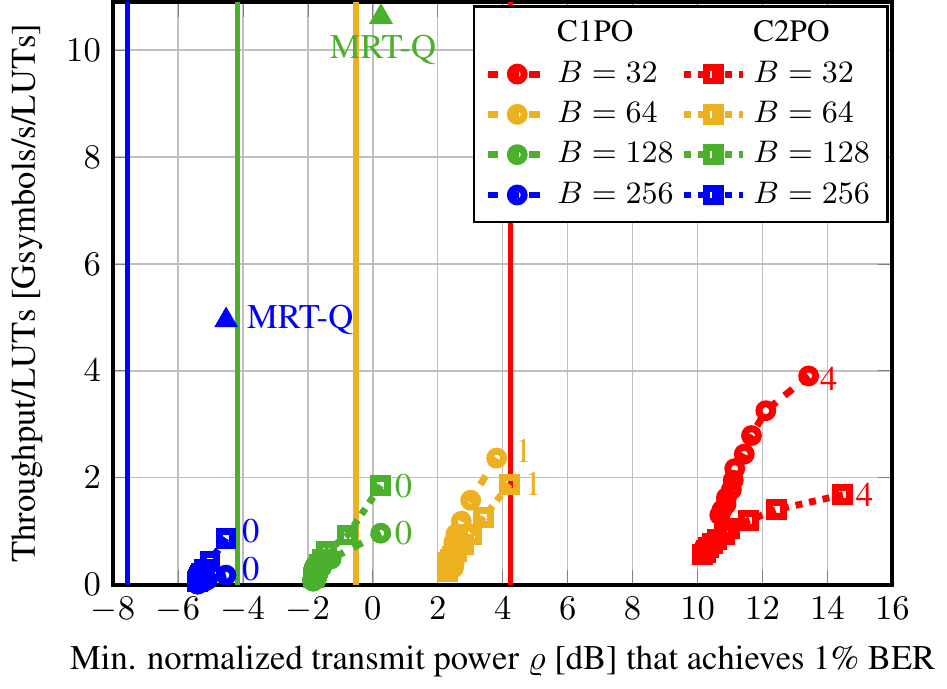}
\caption{Performance--complexity trade-offs for C1PO and C2PO. The numbers next to the curves correspond to the number of iterations $t_\text{max}$. For \mbox{$t_\text{max}=0$}, we directly take the outputs from the initialization step $\bmx^{(1)}=\bH^H\bms$, which is an approach equivalent to {MRT-Q}. The vertical lines show the performance of ZF precoding with infinite-precision DACs. C1PO outperforms C2PO for small BS antenna arrays ($B=32$ and $B=64$); C2PO outperforms C1PO for large antenna arrays  ($B=128$ and $B=256$). {MRT-Q} achieves higher throughput per LUT at the cost of rather poor  performance.}
\label{fig:performancecomplexityradeoff}
\end{figure}

In \fref{fig:performancecomplexityradeoff}, we provide the performance--complexity trade-offs between C1PO (dashed lines with circle markers) and C2PO (dotted lines with square markers) for various BS antenna array sizes. This trade-off is characterized in terms of the minimum normalized transmit power~$\snr$ required to achieve 1\% uncoded BER for BPSK (as in~\fref{fig:BER_32x16_BPSK}); the {hardware efficiency} is characterized by the throughput per area (in terms of billion symbols per second per FPGA LUT). As a reference, we also show the performance for ZF precoding with infinite-precision DACs (vertical lines). As in \fref{fig:BER_ALL}, we consider a transmission to $U=16$ UEs.
{
\fref{fig:performancecomplexityradeoff} shows that, for scenarios with a high normalized transmit power~$\snr$, only a few iterations of our algorithms are required to meet  1\% uncoded BER. As the value of $\snr$ decreases, more iterations are needed, which reduces the throughput and, hence, the hardware efficiency of the circuit.}
We see that for small antenna arrays (i.e., for $B=32$ and $B=64$), C1PO outperforms C2PO, while for large antenna arrays (i.e., for $B=128$ and $B=256$), C2PO significantly outperforms C1PO. We note, however, that the reported hardware efficiency does not take into account the fact that the preprocessing complexity of C1PO would be substantially higher than that of C2PO; see our discussion in \fref{sec:fpgaresults}.
We also observe that only a small number of iterations are required (e.g., $2$~to $4$ iterations) for such large BS antenna arrays to achieve the {error-rate} performance limits of our algorithms.

In \fref{fig:performancecomplexityradeoff}, we additionally show the trade-off achieved by the {MRT-Q} baseline design reported in \fref{sec:MRCbaselinedesign}. Clearly, {MRT-Q} achieves higher throughput per LUT than C1PO and C2PO for large BS arrays ($B=128$ and $B=256$); this gain comes, however, at the cost of rather poor error-rate performance. For small BS antenna arrays  ($B=32$ and $B=64$), {MRT-Q} is unable to achieve the target BER of $1$\%. Hence, {MRT-Q} is only suitable for massive MU-MIMO systems with very high BS-to-UE-antenna ratios in which best-in-class error-rate performance is not the main design objective.

\begin{rem}
{The latency of C1PO and C2PO could be reduced by modifying the architectures proposed in \fref{sec:architecture1} and \fref{sec:architecture2}.
While the proposed architectures only pass one element of the vector $\bmx^{(t)}$ (for C1PO) and one element of the sub-vector $\tilde{\bmx}_w^{(t)}$ (for C2PO) per clock cycle, both architectures could process two or more elements per clock cycle.
Such an approach would significantly decrease the latency and improve the throughput at the cost of increased silicon area.}
\end{rem}


\section{Conclusions}
\label{sec:conclusions}

\sloppy

We have proposed two nonlinear precoding algorithms, namely C1PO and C2PO, which achieve excellent error-rate performance in 1-bit massive MU-MIMO systems at low computational complexity. 
To substantiate this claim, we have designed corresponding VLSI architectures---to the best of our knowledge, the first for 1-bit {precoding in the downlink of} massive MU-MIMO systems---and we have presented FPGA reference implementations for a variety of BS antenna array configurations. 
Our results demonstrate that nonlinear precoding for 1-bit massive MU-MIMO systems is feasible from a hardware implementation perspective, even for antenna arrays with hundreds of BS antennas. As a result, our hardware designs {pave the way for enabling} BS antenna arrays with 1-bit DACs to reliably transmit high-rate data to multiple UEs, which has the potential to {keep hardware} complexity, system costs, and circuit power consumption within manageable limits.

There are many avenues for future work. 
Besides the proposed convergence results, a theoretical error-rate performance analysis of C1PO and C2PO is a challenging open research topic.
Implementing precoders for other nonlinear algorithms, such as SQUID~\cite{jacobsson16c}, which perform better than C1PO and C2PO at low {normalized transmit power $\snr$}, is left for future work.
{The study of 1-bit nonlinear precoders using more realistic system models and a comprehensive cost, power, and performance analysis are interesting research directions.}
Specifically, the design of 1-bit precoding algorithms and hardware accelerators for wideband massive MU-MIMO systems that use orthogonal-frequency division multiplexing (OFDM) is the subject of ongoing work; preliminary results are reported in \cite{jacobsson17globecom}.

\fussy

\appendices

\section{Proof of \fref{thm:c1po}}
\label{app:c1po}

Let $E(\bmz,\bmx) = \vecnorm{\bA\vecz}^2_2  + \gamma\|\bmz-\bmx\|_2^2 - \delta\|\bmx\|_2^2$ denote the objective (\BCRp) minimized by C1PO.
Because~$\setB^B$ is bounded, the sequence of iterates $\{(\bmz^{(t)},\bmx^{(t)})\}$ remains bounded and thus contains a convergent sub-sequence. Denote the limit of this sub-sequence by $(\bmz^{\star},\bmx^{\star})$ and set $E^\star= E(\bmz^{\star},\bmx^{\star}).$  Consider the point $\hat \bmz^{\star} = \argmin_\bmz E(\bmz,\bmx^\star) = (\bI_B + \gamma^{-1}\bA^H\bA)^{-1} \bmx^{\star}.$  If $\hat \bmz^{\star}\neq \bmz^{\star},$ then we have the strict inequality
$$ E( (\hat \bmz^{\star}+ \bmz^{\star})/2,  \bmx^{\star}) <  \frac{1}{2} E( \hat \bmz^{\star},  \bmx^{\star}) +  \frac{1}{2} E( \bmz^{\star},  \bmx^{\star})  = E^\star $$
because $E$ is strongly convex in $\bmz.$  However, this contradicts the fact that $\hat \bmz^{\star} = \argmin_\bmz E(\bmz,\bmx^\star),$ and so it must be the case that $\hat \bmz^{\star} = \bmz^{\star}.$  Because $\delta < \gamma,$ $E$ is strongly convex in $\bmx,$ and a similar argument shows that $\bmx^\star = \argmin_{\bmx\in\setB^B} E(\bmz^\star,\bmx).$  Hence, $(\bmz^{\star},\bmx^{\star})$ minimizes $E$ with respect to $\bmz$ and~$\bmx$ separately; this, combined with the fact that $E$ is differentiable, and $\setB$ coordinate-wise separable, guarantees that $(\bmz^{\star},\bmx^{\star})$ satisfies the first-order conditions for  \text{(\BCRp)}; see Theorem 2 in \cite{tseng2001convergence} and similar arguments in \cite{richtarik2014iteration}.

\section{Proof of \fref{thm:c2po}}
\label{app:c2po}

Let $E(\bmz,\bmx) = \vecnorm{\bA\vecx}^2_2 - \delta\|\bmx\|_2^2$ denote the objective \eqref{eq:directform} minimized by C2PO.
Let $f$ and $g$ be defined as in \eqref{eq:directsplit}.  
Using the definition of the proximal operator \eqref{eq:proxg} together with \eqref{eq:faststep2}, the second update~\eqref{eq:faststep1} of C2PO can be written as
\begin{align*}
 \bmx^{(t+1)}& =  \argmin_{\bmx} g(\bmx) + \frac{1}{2\tau}\|\bmx - (\bmx^{(t)}-\tau \nabla f(\bmx^{(t)}))\|^2 \nonumber\\
  &= \argmin_{\bmx} g(\bmx) +  f(\bmx^{(t)}) + \langle \bmx-\bmx^{(t)},\nabla f(\bmx^{(t)}) \rangle \\
   & \qquad \qquad  \,\,\, + \frac{1}{2\tau}\|\bmx - \bmx^{(t)}\|^2.
\end{align*}
Observe that, whenever $\tau < \|\bA^T\bA\|^{-1}_{2,2},$ the inequality 
$$f(\bmx) \le f(\bmx^{(t)}) + \langle \bmx-\bmx^{(t)},\nabla f(\bmx^{(t)}) \rangle + \frac{1}{2\tau}\|\bmx - \bmx^{(t)}\|^2$$ holds for all $\bmx.$  Using this observation, we can write
\begin{align*}
E(\bmx^{(t+1)})  = &\,  g(\bmx^{(t+1)}) + f(\bmx^{(t+1)})\\ 
   \le &\, g(\bmx^{(t+1)}) +  f(\bmx^{(t)})  +\langle \bmx^{(t+1)}-\bmx^{(t)},\nabla f(\bmx^{(t)}) \rangle \\
    & + \frac{1}{2\tau}\|\bmx^{(t+1)} - \bmx^{(t)}\|^2\\
    =&\,  \min_{\bmx} \, g(\bmx) +  f(\bmx^{(t)}) + \langle \bmx-\bmx^{(t)},\nabla f(\bmx^{(t)}) \rangle \\
    & \qquad + \frac{1}{2\tau}\|\bmx - \bmx^{(t)}\|^2 \\
  \le &\, g(\bmx^{(t)}) +  f(\bmx^{(t)}) = E( \bmx^{(t)}).
\end{align*}
This shows that the sequence $\{E(\bmx^{(t)})\}$ is monotonically decreasing.  Since the sequence is bounded below, there is some limit $L = \lim_{t\to\infty} E(\bmx^{(t)}).$
Let $\{\bmx^{(t_{k})}\}$ be a convergent sub-sequence of iterates (which must exist because the iterates are bounded) with limit point $\bmx^{\star}$. Let
\begin{align}
\bar \bmx^{\star}= \argmin_\bmx \,\, & g(\bmx) +  f(\bmx^{\star}) + \langle \bmx-\bmx^{\star},\nabla f(\bmx^{\star}) \rangle \notag \\
& + \frac{1}{2\tau}\|\bmx - \bmx^{\star}\|^2 \label{eq:lastStepOfProof}
\end{align}
be the result of applying the C2PO iteration starting at $\bmx^{\star}$.
 Observing that $E(\bmx^{(t_{k}+1)})\le E(\bmx^{(t_{k})}) \le E(\bmx^{(t_{k}-1)}),$ and letting $k\to\infty,$ we find that  $E(\bar \bmx^{\star})=E(\bmx^{\star})=L,$ and so $\bmx^{\star}$ is a minimizer of  \eqref{eq:lastStepOfProof}.  This is only possible if $0\in \partial g(\bmx^{\star}) + \nabla f(\bmx^{\star}),$ in which case $\bmx^{\star}$ is a stationary point.
 

\section*{Acknowledgments}

The authors would like to thank O. Tirkkonen for insightful discussions on 1-bit precoding. The authors also thank A. Burg for discussions on the hardware architecture and R. Manohar for pointing us to its connection to Cannon's algorithm.
The work of O.~Casta\~neda and C.~Studer was supported in part by Xilinx, Inc.\ and by the US National Science Foundation (NSF) under grants ECCS-1408006, CCF-1535897,  CAREER CCF-1652065, and CNS-1717559. The work of S.~Jacobsson and G.~Durisi was supported by the Swedish Foundation for Strategic Research under grant ID14-0022, and by the Swedish Governmental Agency for Innovation Systems (VINNOVA) within the center ChaseOn. The work of T.~Goldstein was supported in part by the US NSF under grant CCF-1535902 and by the US Office of Naval Research under grant N00014-17-1-2078.



\balance


\begin{thebibliography}{10}
\providecommand{\url}[1]{#1}
\csname url@samestyle\endcsname
\providecommand{\newblock}{\relax}
\providecommand{\bibinfo}[2]{#2}
\providecommand{\BIBentrySTDinterwordspacing}{\spaceskip=0pt\relax}
\providecommand{\BIBentryALTinterwordstretchfactor}{4}
\providecommand{\BIBentryALTinterwordspacing}{\spaceskip=\fontdimen2\font plus
\BIBentryALTinterwordstretchfactor\fontdimen3\font minus
  \fontdimen4\font\relax}
\providecommand{\BIBforeignlanguage}[2]{{%
\expandafter\ifx\csname l@#1\endcsname\relax
\typeout{** WARNING: IEEEtran.bst: No hyphenation pattern has been}%
\typeout{** loaded for the language `#1'. Using the pattern for}%
\typeout{** the default language instead.}%
\else
\language=\csname l@#1\endcsname
\fi
#2}}
\providecommand{\BIBdecl}{\relax}
\BIBdecl

\bibitem{castaneda17icassp}
O.~{Casta\~neda}, T.~Goldstein, and C.~Studer, ``{POKEMON:} a non-linear
  beamforming algorithm for 1-bit massive {MIMO},'' in \emph{IEEE Intl. Conf.
  on Acoustics, Speech, and Sig. Proc. (ICASSP)}, New Orleans, LA, Mar. 2017.

\bibitem{rusek14a}
F.~Rusek, D.~Persson, B.~Kiong, E.~G. Larsson, T.~L. Marzetta, O.~Edfors, and
  F.~Tufvesson, ``Scaling up {MIMO}: Oppurtunities and challenges with very
  large large arrays,'' \emph{{IEEE} Signal Process. Mag.}, vol.~30, no.~1, pp.
  40--60, Jan. 2013.

\bibitem{larsson14a}
E.~G. Larsson, F.~Tufvesson, O.~Edfors, and T.~L. Marzetta, ``Massive {MIMO}
  for next generation wireless systems,'' \emph{{IEEE} Commun. Mag.}, vol.~52,
  no.~2, pp. 186--195, Feb. 2014.

\bibitem{lu14a}
L.~Lu, G.~Ye~Li, A.~L. Swindlehurst, A.~Ashikhmin, and R.~Zhang, ``An overview
  of massive {MIMO}: Benefits and challenges,'' \emph{{IEEE} J. Sel. Topics
  Signal Process.}, vol.~8, no.~5, pp. 742--758, Oct. 2014.

\bibitem{li16globalsip}
K.~Li, R.~Sharan, Y.~Chen, T.~Goldstein, J.~R. Cavallaro, and C.~Studer,
  ``Decentralized beamforming for massive {MU-MIMO} on a {GPU} cluster,'' in
  \emph{4th IEEE Global Conf. on Sig. and Info. Proc. (GlobalSIP)}, Washington,
  D.C., Dec. 2016.

\bibitem{li2017decentralized}
\BIBentryALTinterwordspacing
------, ``Decentralized baseband processing for massive {MU-MIMO} systems,''
  Feb. 2017. [Online]. Available: \url{arXiv: 1702.04458}
\BIBentrySTDinterwordspacing

\bibitem{jacobsson16c}
S.~Jacobsson, G.~Durisi, M.~Coldrey, T.~Goldstein, and C.~Studer, ``Quantized
  precoding for massive {MU-MIMO},'' \emph{IEEE Trans. Comm.; arXiv preprint:
  1610.07564}, Jul. 2016.

\bibitem{jacobsson16b}
------, ``Nonlinear 1-bit precoding for massive {MU-MIMO} with higher-order
  modulation,'' in \emph{Proc. Asilomar Conf. Signals, Syst., Comput.}, Pacific
  Grove, CA, Nov. 2016, pp. 763--767.

\bibitem{jedda16a}
H.~Jedda, J.~A. Nossek, and A.~Mezghani, ``Minimum {BER} precoding in 1-bit
  massive {MIMO} systems,'' in \emph{{IEEE} Sensor Array and Multichannel Sig.
  Proc. Workshop (SAM)}, Rio de Janeiro, Brazil, Jul. 2016.

\bibitem{tirkkonen17a}
O.~Tirkkonen and C.~Studer, ``Subset-codebook precoding for 1-bit massive
  multiuser {MIMO},'' in \emph{Conf. on Info. Sciences and Systems (CISS)},
  Baltimore, MA, Mar. 2017.

\bibitem{mezghani09c}
A.~Mezghani, R.~Ghiat, and J.~A. Nossek, ``Transmit processing with low
  resolution {D/A}-converters,'' in \emph{Proc. IEEE Int. Conf. Electron.,
  Circuits, Syst. (ICECS)}, Yasmine Hammamet, Tunisia, Dec. 2009, pp. 683--686.

\bibitem{saxena16a}
A.~K. Saxena, I.~Fijalkow, and A.~L. Swindlehurst, ``On one-bit quantized {ZF}
  precoding for the multiuser massive {MIMO} downlink,'' in \emph{{IEEE} Sensor
  Array and Multichannel Sig. Proc. Workshop (SAM)}, Rio de Janeiro, Brazil,
  Jul. 2016.

\bibitem{guerreiro16a}
R.~D. J.~Guerreiro and P.~Montezuma, ``Use of 1-bit digital-to-analogue
  converters in massive {MIMO} systems,'' \emph{{IEEE} Electron. Lett.},
  vol.~52, no.~9, pp. 778--779, Apr. 2016.

\bibitem{usman16a}
O.~B. Usman, H.~Jedda, A.~Mezghani, and J.~A. Nossek, ``{MMSE} precoder for
  massive {MIMO} using 1-bit quantization,'' in \emph{Proc. IEEE Int. Conf.
  Acoust., Speech, Signal Process. (ICASSP)}, Shanghai, China, Mar. 2016, pp.
  3381--3385.

\bibitem{shah2016biconvex}
S.~Shah, A.~K. Yadav, C.~D. Castillo, D.~W. Jacobs, C.~Studer, and
  T.~Goldstein, ``Biconvex relaxation for semidefinite programming in computer
  vision,'' in \emph{European Conf. on Comp. Vision (ECCV)}.\hskip 1em plus
  0.5em minus 0.4em\relax Springer, Sep. 2016, pp. 717--735.

\bibitem{risi14a}
\BIBentryALTinterwordspacing
C.~Risi, D.~Persson, and E.~G. Larsson, ``Massive {MIMO} with 1-bit {ADC},''
  Apr. 2014. [Online]. Available: \url{http://arxiv.org/abs/1404.7736}
\BIBentrySTDinterwordspacing

\bibitem{jacobsson15a}
S.~Jacobsson, G.~Durisi, M.~Coldrey, U.~Gustavsson, and C.~Studer, ``One-bit
  massive {MIMO}: Channel estimation and high-order modulations,'' in
  \emph{Proc. IEEE Int. Conf. Commun. Workshop (ICCW)}, London, U.K., June
  2015, pp. 1304--1309.

\bibitem{li16b}
Y.~Li, C.~Tao, G.~Seco-Granados, A.~Mezghani, A.~L. Swindlehurst, and L.~Liu,
  ``Channel estimation and performance analysis of one-bit massive {MIMO}
  systems,'' \emph{IEEE Trans. Signal Process.}, vol.~65, no.~15, pp.
  4075--4089, May 2016.

\bibitem{mollen16c}
C.~Moll{\'e}n, J.~Choi, E.~G. Larsson, and R.~W. {Heath Jr.}, ``Uplink
  performance of the wideband massive uplink {MIMO} with one-bit {ADCs},''
  \emph{IEEE Trans. Wireless Commun.}, vol.~16, no.~1, pp. 87--100, 2017.

\bibitem{studer15a}
C.~Studer and G.~Durisi, ``Quantized massive {MU-MIMO-OFDM} uplink,''
  \emph{{IEEE} Trans. Commun.}, vol.~64, no.~6, pp. 2387--2399, Jun. 2016.

\bibitem{WYWDCS2014}
M.~Wu, B.~Yin, G.~Wang, C.~Dick, J.~Cavallaro, and C.~Studer, ``Large-scale
  {MIMO} detection for {3GPP LTE}: Algorithm and {FPGA} implementation,''
  \emph{{IEEE} J. Sel. Topics Signal Process.}, vol.~8, no.~5, pp. 916--929,
  Oct. 2014.

\bibitem{wu2016efficient}
Z.~Wu, C.~Zhang, Y.~Xue, S.~Xu, and X.~You, ``Efficient architecture for
  soft-output massive {MIMO} detection with {Gauss-Seidel} method,'' in
  \emph{IEEE Int. Symp. on Circuits and Systems (ISCAS)}, Montreal, Canada,
  Aug. 2016, pp. 1886--1889.

\bibitem{wu2016high}
M.~Wu, C.~Dick, J.~R. Cavallaro, and C.~Studer, ``High-throughput data
  detection for massive {MU-MIMO-OFDM} using coordinate descent,'' \emph{IEEE
  Trans. on Circuits and Systems I: Regular Papers}, vol.~63, no.~12, pp.
  2357--2367, Nov. 2016.

\bibitem{castaneda2016data}
O.~Casta{\~n}eda, T.~Goldstein, and C.~Studer, ``Data detection in large
  multi-antenna wireless systems via approximate semidefinite relaxation,''
  \emph{IEEE Trans. on Circuits and Systems I: Regular Papers}, vol.~63,
  no.~12, pp. 2334--2346, Nov. 2016.

\bibitem{barrenechea2010design}
M.~Barrenechea, L.~Barbero, M.~Mendicute, and J.~Thompson, ``Design and
  hardware implementation of a low-complexity multiuser vector precoder,'' in
  \emph{Conf. on Design and Architectures for Sig. and Image Proc. (DASIP)},
  Oct. 2010, pp. 160--167.

\bibitem{prabhu2014hardware}
H.~Prabhu, O.~Edfors, J.~Rodrigues, L.~Liu, and F.~Rusek, ``Hardware efficient
  approximative matrix inversion for linear pre-coding in massive {MIMO},'' in
  \emph{IEEE Intl. Symp. on Circuits and Systems (ISCAS)}, June 2014, pp.
  1700--1703.

\bibitem{shepard2013practical}
C.~Shepard, N.~Anand, and L.~Zhong, ``Practical performance of {MU-MIMO}
  precoding in many-antenna base stations,'' in \emph{Proc. of the 2013
  workshop on Cellular networks: operations, challenges, and future
  design}.\hskip 1em plus 0.5em minus 0.4em\relax ACM, June 2013, pp. 13--18.

\bibitem{prabhu2017isscc}
H.~Prabhu, O.~Edfors, J.~Rodrigues, L.~Liu, and F.~Rusek, ``A 60~p{J}/b
  300~{M}b/s 128$\times$8 massive {MIMO} precoder-detector in 28nm {FD-SOI},''
  in \emph{IEEE Intl. Solid-State Circuits Conf. (ISSCC)}, San Francisco,
  United States of America, Feb. 2017, pp. 60--61.

\bibitem{bjornson14b}
E.~Bj{\"o}rnson, M.~Bengtsson, and B.~Ottersten, ``Optimal multiuser transmit
  beamforming: A difficult problem with a simple solution structure,''
  \emph{{IEEE} Signal Process. Mag.}, vol.~31, no.~4, pp. 142--148, Jul. 2014.

\bibitem{bjornson13a}
E.~Bj{\"o}rnson and E.~Jorswieck, ``Optimal resource allocation in coordinated
  multi-cell systems,'' \emph{Foundations and Trends in Communications and
  Information Theory}, vol.~9, no. 2-3, pp. 113--381, 2013.

\bibitem{joham05a}
M.~Joham, W.~Utschick, and J.~A. Nossek, ``Linear transmit processing in {MIMO}
  communications systems,'' \emph{{IEEE} Trans. Signal Process.}, vol.~53,
  no.~8, pp. 2700--2712, Aug. 2005.

\bibitem{shi07a}
S.~Shi, M.~Schubert, and H.~Boche, ``Downlink {MMSE} transceiver optimization
  for multiuser {MIMO} systems: Duality and sum-{MSE} minimization,''
  \emph{{IEEE} Trans. Signal Process.}, vol.~55, no.~11, pp. 5436--5446, Nov.
  2007.

\bibitem{agrell02a}
E.~Agrell, T.~Eriksson, A.~Vardy, and K.~Zeger, ``Closest point search in
  lattices,'' \emph{{IEEE} Trans. Inf. Theory}, vol.~48, no.~8, pp. 2201--2214,
  Aug. 2002.

\bibitem{fincke85a}
U.~Fincke and M.~Pohst, ``Improved methods for calculating vectors of short
  length in a lattice, including a complexity analysis,'' \emph{Math. Comput.},
  vol.~44, no. 170, pp. 463--471, Apr. 1985.

\bibitem{verdu89a}
S.~Verd{\'u}, ``Computational complexity of multiuser detection,''
  \emph{Algorithmica}, vol.~4, no.~1, pp. 303--312, 1989.

\bibitem{GSB14}
\BIBentryALTinterwordspacing
T.~Goldstein, C.~Studer, and R.~G. Baraniuk, ``A field guide to
  forward-backward splitting with a {FASTA} implementation,'' Nov. 2014.
  [Online]. Available: \url{http://arxiv.org/abs/1411.3406}
\BIBentrySTDinterwordspacing

\bibitem{BT09}
A.~Beck and M.~Teboulle, ``A fast iterative shrinkage-thresholding algorithm
  for linear inverse problems,'' \emph{SIAM J. Imag. Sci.}, vol.~2, no.~1, pp.
  183--202, Jan. 2009.

\bibitem{goldstein2010high}
T.~Goldstein and S.~Setzer, ``High-order methods for basis pursuit,''
  \emph{UCLA CAM Report}, pp. 10--41, 2010.

\bibitem{parikh2014proximal}
N.~Parikh and S.~Boyd, ``Proximal algorithms,'' \emph{Foundations and
  Trends{\textregistered} in Optimization}, vol.~1, no.~3, pp. 127--239, Jan.
  2014.

\bibitem{GV96}
G.~H. Golub and C.~F. {van Loan}, \emph{Matrix Computations}, 3rd~ed.\hskip 1em
  plus 0.5em minus 0.4em\relax The Johns Hopkins Univ. Press, 1996.

\bibitem{CannonThesis}
L.~Cannon, ``A cellular computer to implement the {Kalman} filter algorithm,''
  Ph.D. dissertation, Montana State University, United States, 1969.

\bibitem{jacobsson17globecom}
S.~Jacobsson, G.~Durisi, M.~Coldrey, and C.~Studer, ``Massive {MU-MIMO-OFDM}
  downlink with one-bit {DACs} and linear precoding,'' in \emph{Proc. IEEE
  Global Telecommun. Conf. (GLOBECOM)}, Singapore, Dec. 2017.

\bibitem{tseng2001convergence}
P.~Tseng, ``Convergence of a block coordinate descent method for
  nondifferentiable minimization,'' \emph{J. of Opt. Theory and Applications},
  vol. 109, no.~3, pp. 475--494, June 2001.

\bibitem{richtarik2014iteration}
P.~Richt{\'a}rik and M.~Tak{\'a}{\v{c}}, ``Iteration complexity of randomized
  block-coordinate descent methods for minimizing a composite function,''
  \emph{Mathematical Programming}, vol. 144, no. 1-2, pp. 1--38, 2014.

\end{thebibliography}
\end{document}